\documentclass[superscriptaddress,nofootinbib,a4paper,11pt]{article}
\usepackage{jheppub}
\usepackage{graphicx}
\usepackage{subcaption}
\usepackage{float}
\usepackage{amsmath}
\usepackage{amssymb}
\usepackage[utf8]{inputenc}
\usepackage{hyperref}
\usepackage{color}
\usepackage{slashed}
\allowdisplaybreaks

\newcommand{\be}{\begin{equation}}
\newcommand{\ee}{\end{equation}}

\begin{document}

\title{\boldmath Investigating the collinear splitting effects of boosted dark matter at neutrino detectors}

\author[a]{Jinmian Li,}
\author[b,c]{Junle Pei,}
\author[a]{Cong Zhang}

\affiliation[a]{College of Physics, Sichuan University, Chengdu 610065, China}
\affiliation[b]{Institute of High Energy Physics, Chinese Academy of Sciences, Beijing 100049, China}
\affiliation[c]{Spallation Neutron Source Science Center, Dongguan 523803, China}

\emailAdd{jmli@scu.edu.cn}
\emailAdd{peijunle@ihep.ac.cn}
\emailAdd{zhangcong.phy@gmail.com}

\abstract{
We study the probing prospects of cosmic ray boosted dark matter (DM) in the framework of simplified electron-philic dark photon model. 
Focusing on the dark matter and dark photon masses around keV $\sim $ MeV scale, we consider the bounds obtained from the XENON1T and Super-K experiments. 
The electron bound state effects are treated carefully in calculating the XENON1T constraint. 
As for the detection at neutrino detector where the energy threshold is relatively higher, the large logarithmic effects induced by the scale hierarchy between the masses and momentum transfer are considered by introducing the DM parton distribution function (PDF). 
The logarithmic effects will reduce the electron recoil rate for DM scattering in neutrino detectors. 
Moreover, we find the DUNE and JUNO experiments provide high sensitivities for probing the dark photon component in the DM PDF through the dark Compton process. 
We also check the Bullet Cluster constraint on the DM self-scattering cross section.
}


\maketitle

\section{Introduction}\label{sec:intro} 

Dark Matter (DM) is one of the most evident new physics beyond the Standard Model (SM). 
The potential DM signals have been searched in various forms in terms of DM direct detection, indirect detection and collider searches. 
However, so far these observations which concern the particle properties of DM remain null about the sign of DM.  
The DM direct detection experiments utilizing the signal of recoiled nucleon or electron from the DM scattering aim to search for the DM in the Galactic halo with characteristic velocities of $v_{\chi} \sim 10^{{-3}} c$. 
Due to the smallness of the kinetic energy of halo DM, those analyses are most sensitive when the DM mass is comparable to the target mass. They will lose the sensitivities rapidly for DM mass below GeV scale~\cite{MarrodanUndagoitia:2015veg}. 

Recently, it has been pointed out in literature that the halo DM can be boosted in several ways in astrophysics, such as primordial black holes evaporation~\cite{Carr:2020gox}, two-component DM annihilation~\cite{Agashe:2014yua}, DM semi-annihilation~\cite{DEramo:2010keq} and so on~\cite{Jaeckel:2020oet,Herrera:2021puj}. 
This sub-dominant and (near-)relativistic component of DM can produce a detectable signal at direct detection experiment even for DM masses well below 1 GeV. It could be a smoking gun for the DM discovery. 
In particular, in DM models with freezeout mechanism, the interactions between the DM and the SM particles induce the scattering between the DM and high energy cosmic ray (CR) particles. 
The up-scattering of DM particles by CR will inevitably give rise to a non-negligible flux of boosted DM in detect detection experiment~\cite{Bringmann:2018cvk,Cappiello:2018hsu,Yin:2018yjn,Flambaum:2020xxo,PandaX-II:2021kai,Xia:2021vbz,Bramante:2021dyx,Jho:2021rmn,Feng:2021hyz,Chen:2021ifo,Elor:2021swj,Xia:2022tid,Bardhan:2022ywd,Alvey:2022pad}. 
Even very light DM particles can deposit large energy at the target detector in this case~\cite{Ge:2020yuf,Lei:2020mii,Dent:2019krz,Cho:2020mnc,Wang:2021nbf}. 
It has been studied that electronic recoil data at XENON1T detector~\cite{XENON:2019gfn,XENON:2020rca} can be sensitive to the DM with mass around MeV scale~\cite{Wang:2019jtk,Cao:2020bwd,Jho:2020sku,Emken:2021vmf,Das:2021lcr,Xia:2020apm}. 

Moreover, large-volume neutrino experiments installed deep underground such as Super-K~\cite{Super-Kamiokande:2002weg}, DUNE~\cite{DUNE:2020ypp} and JUNO~\cite{JUNO:2021vlw} have been considered to probe the boosted DM flux as well. 
Compared with the DM direct detection, the energy threshold for the signal electron is much higher in those neutrino detectors, which will reject a large amount of DM scattering events. On the other hand, the statistic at neutrino detectors can be regained via the considerably larger size (typically with several kilotons of materials).
As a result, the sensitivities of the neutrino detector to the DM models can be complementary to those of the DM direct detections~\cite{Ema:2018bih,Cappiello:2019qsw,Berger:2019ttc,Kim:2020ipj,Guo:2020drq,DeRoeck:2020ntj,Harnik:2020ugb,Ema:2020ulo,Dent:2020syp,Granelli:2022ysi,Berger:2022cab}. 
In such scenarios, the energy scale of the DM-SM particle scattering is usually much higher than the mass scale of the DM sector. 
As has been well known in the SM, the hierarchy between two scales in a process will lead to large logarithms~\cite{PhysRevLett.84.4810,Ciafaloni:2003xf,Ciafaloni:2005fm,Han:2020uid}. 
Those large logarithms will give significant contributions to the differential cross section of DM scattering.
They need to be resumed using the {\it Dokshitzer-Gribov-Lipatov-Altarelli-Parisi} (DGLAP) equations~\cite{DGLAP1,DGLAP2,DGLAP3,DGLAP4}. 
The corresponding scattering cross section can be calculated following the  factorization theorem~\cite{Han:2020uid}.

In this work, based on a simplified version of electron-philic dark photon model with fermionic DM and focusing on the CR up-scattering DM acceleration mechanism, we consider the boosted DM - bounded electron scattering process at the XENON1T detector as well as several neutrino detectors, including Super-K, DUNE and JUNO. 
In particular, in calculating the electron recoil rate at the XENON1T, we carefully treat the bound state effects of the initial electron. 
As for the signals at neutrino detectors, we first calculate the DM parton distribution function (PDF) by using the DGLAP equations, and consider the scattering rate between all components in the DM PDF and the electron. 
The high energy electron recoil signals can be induced by DM, anti-DM and dark photon components, which are stringently constrained by the Super-K data (161.9 kiloton-years exposure)~\cite{Super-Kamiokande:2017dch}.
Furthermore, we find the dark photon component in the DM PDF can induce the dark Compton scattering~\cite{Hochberg:2021zrf}, giving the mono-energetic photon in the final states. 
Although the Super-K detector which is water Cherenkov based is not able to probe such a monochromatic photon discriminately from electron, future neutrino detectors such as DUNE and JUNO can probe such photons efficiently, as have been studied in Ref.~\cite{Cui:2022owf}. 
We explicitly calculate the rate for photon signal at those two detectors, and find that this signal provides better sensitivity than the electron recoil signal in the region with relatively light DM and heavy dark photon.
The corresponding bounds on DM self-scattering cross section from Bullet Cluster~\cite{Markevitch:2003at,Clowe:2003tk,Randall:2008ppe} are investigated over the parameter regions of interest.

This paper is organized as follows. 
In section~\ref{sec2}, we calculate the DM PDF by solving the DGLAP equations in the simplified dark photon model. 
In section~\ref{sec3}, we estimate the flux of boosted DM that is induced by the cosmic ray up-scattering. 
The differential recoil rates for DM direct detection experiments and neutrino detectors are calculated in section~\ref{sec4}. 
The corresponding results are shown and discussed in section~\ref{sec5}. 
We conclude our study in section~\ref{sec6}. 
Moreover, the paper is supplemented with appendices~\ref{appendix-direct} and~\ref{appendix-sk}, to provide technical details in calculating the recoil rates for both DM direct detections and neutrino detectors. 

\section{The dark matter parton distribution function}\label{sec2}

Considering a spacelike branching process $A\rightarrow B+C$, the four-momentum of particles is chosen as 
\begin{align}
&P_A=(E_A,~0,~0,~E_A-\frac{m_A^2}{2E_A})~,\\
&P_B=(zE_A,~k_T,~0,~zE_A+\frac{k_T^2-\bar{z}m_A^2+m_C^2}{2\bar{z}E_A})~,\\
&P_C=(\bar{z}E_A,~-k_T,~0,~\bar{z}E_A-\frac{k_T^2+m_C^2}{2\bar{z}E_A})~,
\end{align}
where $z$ ranges in $(0,~1)$, $\bar{z}=1-z$, and $E_A^2\gg k_T^2,m_{i}^2~(i=A,B,C)$. By ignoring the terms proportional to $ \frac{k_T^2~\text{or}~m_i^2}{E_A^2}~(i=A,B,C) $, we can obtain
\begin{align}
&P_A^2=m_A^2~,\\
&P_B^2=-\frac{k_T^2+zm_C^2-z\bar{z}m_A^2}{\bar{z}},\\
&P_C^2=m_C^2~,
\end{align}
which means A and C particles are on-shell but B particle is off-shell under the limit of $E_A^2\gg k_T^2,m_{i}^2~(i=A,B,C)$. The corresponding splitting function  can be expressed as 
\begin{align}
\frac{d \mathcal{P}_{A \rightarrow B+C}}{d z d k_{T}^{2}} \simeq \frac{1}{N} \frac{1}{16 \pi^{2}} \frac{z \bar{z}\left|M_{\text{split}}\right|^{2}}{\left(k_{T}^{2}+\bar{z} m_{B}^{2}+z m_{C}^{2}-z \bar{z} m_{A}^{2}\right)^{2}}~,
\end{align}
where the matrix-element square ${\left|M_{\text{split}}\right|^2}$ is computed from the amputated $A\to B+C$ Feynman diagram with on-shell polarization vectors. $N=2~(1)$ is taken when $B$ and $C$ are identical (non-identical) particles. 

In our simplified model with the interaction of 
\begin{align}
\mathcal{L} \supset g^\prime A_\mu^\prime\bar{\chi}\gamma^\mu\chi \label{eq:model1}
\end{align}
between Dirac fermion DM $\chi~({\bar{\chi}})$ and dark photon $A^\prime$, the 
splitting functions of different spacelike branching processes are summarized in Table~\ref{tab:splitf}, where $m_{\chi/A^\prime}$ is the mass of $\chi~(A^\prime)$ particle, $\alpha^\prime=\frac{g^{\prime 2}}{4\pi}$, and $A^\prime_{T/L}$ denotes the dark photon mode of transverse/longitudinal polarization. All the splitting functions in Table~\ref{tab:splitf} have been averaged (summed) over polarizations  of the corresponding initial (final) particles. The splitting function of $\chi / \bar{\chi} \rightarrow \chi / \bar{\chi}+A_{T/L}^{\prime}$ can be obtained by using the relation
\begin{align}
P_{A\rightarrow B+C}(z)=P_{A\rightarrow C+B}(\bar{z})~.
\end{align}
It is noted that terms proportional to $(k_T^2+\bar{z}m_B^2+zm_C^2-z\bar{z}m_A^2)$ in $M_{\text{split}}$ that used to calculate the splitting functions of processes involving the longitudinal mode of dark photon have to be eliminated according to the Reference~\cite{Chen:2016wkt}.

\begin{table}[htb]
	\centering
	\begin{tabular}{cc}  
		\hline
		$A\rightarrow B+C$&   $\frac{d \mathcal{P}_{A \rightarrow B+C}}{d z d k_{T}^{2}}=P_{A \rightarrow B+C}(z)$ \\
		\hline 
		$\chi / \bar{\chi} \rightarrow A_{T}^{\prime}+\chi / \bar{\chi}$ &   $\frac{\alpha^{\prime}}{2 \pi} k_{T}^{2} \frac{\frac{1+\bar{z}^{2}}{z}-\frac{\bar{z}}{z} \frac{2 m_{\chi}^{2} z^{2}+m_{A^\prime}^{2}\left(1+\bar{z}^{2}\right)}{k_{T}^{2}+m_{\chi}^{2} z^{2}+m_{A^\prime}^{2} \bar{z}}}{k_{T}^{2}+m_{\chi}^{2} z^{2}+m_{A^\prime}^{2} \bar{z}}$ \\
		$\chi / \bar{\chi} \rightarrow A_{L}^{\prime}+\chi / \bar{\chi}$ & $\frac{\alpha^{\prime}}{\pi} k_{T}^{2} \frac{m_{A^\prime}^{2} \bar{z}^{2}}{z\left(k_{T}^{2}+m_{\chi}^{2} z^{2}+m_{A^\prime}^{2} \bar{z}\right)^{2}}$  \\
		$A_{T}^{\prime} \rightarrow \bar{\chi} / \chi+\chi / \bar{\chi}$ & $\frac{\alpha^{\prime}}{2 \pi} k_{T}^{2} \frac{z^{2}+\bar{z}^{2}+\frac{z \bar{z}\left(2 m_{\chi}^{2}+m_{A^\prime}^{2}\left(z^{2}+\bar{z}^{2}\right)\right)}{k_{T}^{2}+m_{\chi}^{2}-m_{A^\prime}^{2} z \bar{z}}}{k_{T}^{2}+m_{\chi}^{2}-m_{A^\prime}^{2} z \bar{z}}$ \\
		$A_{L}^{\prime} \rightarrow \bar{\chi} / \chi+\chi / \bar{\chi}$  & $\frac{2 \alpha^{\prime}}{\pi} k_{T}^{2} \frac{m_{A^\prime}^{2} z^{2} \bar{z}^{2}}{\left(k_{T}^{2}+m_{\chi}^{2}-m_{A^\prime}^{2} z \bar{z}\right)^{2}}$ \\
		\hline
	\end{tabular}
	\caption{\label{tab:splitf} Splitting functions involving $\chi$, $\bar{\chi}$, and $A^\prime$.} 
\end{table}

We use $f_i\left({k_{T}}, x\right)$ to denote the PDF of the particle $i$ ($i=\chi,\bar{\chi},A^\prime_T,A^\prime_L$) with an energy fraction $x$ at a factorization scale $k_T$. $f_i(k_T,x)$ evolves according to the DGLAP equations~\footnote{Since $P_{A\rightarrow B+C}(z)$ may diverge at $z=0$ or $1$, the same technologies as those in Reference~\cite{Botje:2010ay} have been used to regularize the possible divergences in our numerical calculations.}
\begin{align}
&\frac{d f_{i}\left(k_T, x\right)}{d \ln k_{T}^{2}}=\sum_{m, n} N\int_{x}^{1} \frac{dz}{z}  P_{m \rightarrow i+n}\left(z \right)  f_{m}\left(k_T, \frac{x}{z}\right) -\sum_{j, k} \int_{0}^{1} d z P_{i \rightarrow j+k}(z) f_{i}\left(k_T, x\right)~,
\end{align}
where $N=2~(1)$ is taken when $n=i$ ($n\ne i$). The initial conditions of the PDFs are
\begin{align}
f_{i}\left(Q_0, x\right)=\begin{cases}
\delta(1-x), ~~~~~i=\chi\\
0,~~~~~~~~~~~~~~i\ne\chi
\end{cases},~~~~~~~~~~~Q_0=\text{max}\left(m_{A^\prime},m_\chi\right) ~.
\end{align}

\begin{figure}[htb]
\centering
\includegraphics[width=0.48\textwidth]{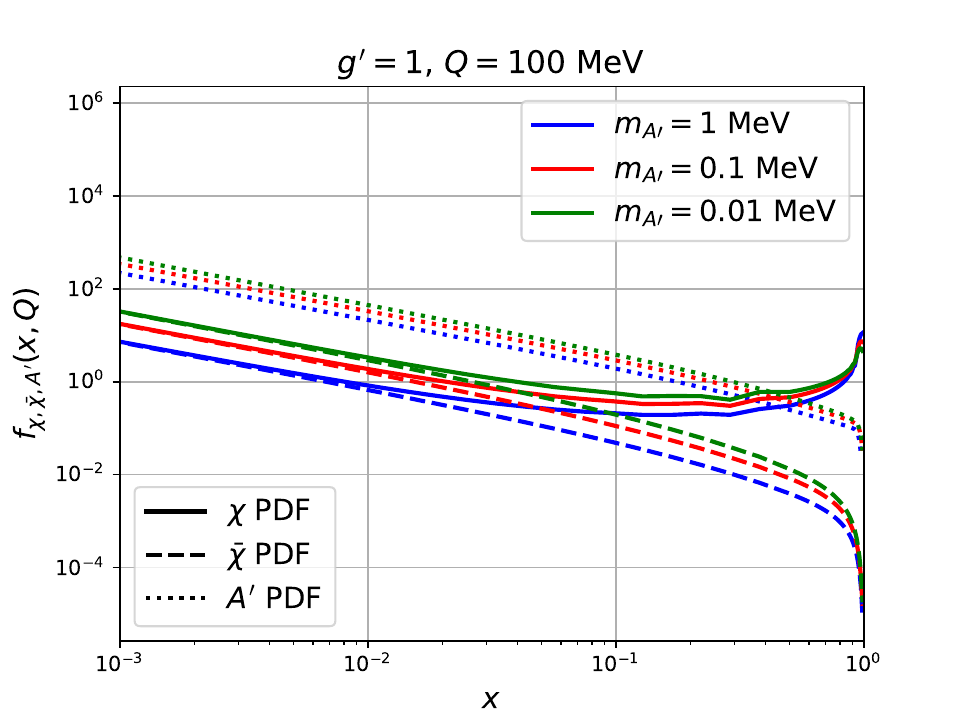}
\includegraphics[width=0.48\textwidth]{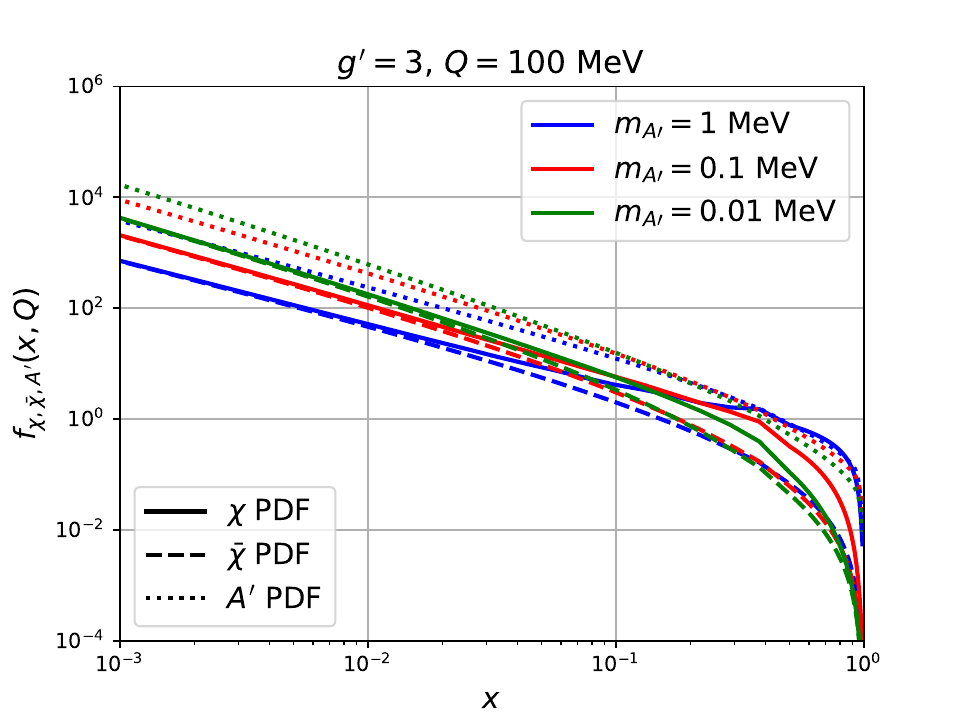}
\caption{The PDFs of $\chi$ (solid line), $\bar{\chi}$ (dashed line) and $A^\prime$ (dotted line) at factorization scale of 100 MeV. 
Different colors indicate the dark photon mass as shown in the legend. 
The coupling $g^\prime=1$ in left panel and  $g^\prime=3$ in right panel. The dark matter mass is taken to be 0.01 MeV.
\label{pdfs}}
\end{figure}

In Figure~\ref{pdfs}, we plot the PDFs for $\chi$, $\bar{\chi}$ and $A^\prime$ with parameter choices that are most relevant to this work as will be discussed later. 
Given the DGLAP evolution equation, at a scale $Q \gg m_{\chi, A^\prime}$, there are large fractions of energy carried by $\bar{\chi}$ and $A^\prime$ in the DM PDF. 
However, we note that, as the PDFs evolve with $\ln k_T^2$, varying the scale within one order of magnitude will not lead to dramatic difference in the PDFs. 
For $g^\prime =1$, the dark photon fraction $f_{A^\prime}$ can exceed the DM fraction $f_{\chi}$ when $x \lesssim 0.5$. The anti-DM fraction $f_{\bar{\chi}}$ becomes comparable to $f_{\chi}$ for $x\lesssim 10^{-2}$. Moreover, lighter $A^\prime$ not only leads to higher fractions for $\bar{\chi}$ and $A^\prime$, but also gives higher $f_{\chi}$ in the low $x$ region. 
As the coupling approaching the perturbative limit $g^\prime =3$, $f_\chi$ no longer has the peak around $x \sim 1$, due to the intensive splittings. And the dark photon fraction even becomes more important than the DM fraction for $m_{A\prime} \lesssim 1$ MeV.  
It is also interesting to observe that in the region $x\gtrsim 0.4$, the dark photon fraction decreases for decreasing $m_{A'}$, as opposed to the low $x$ region. 

\section{Boosted dark matter from cosmic ray acceleration}\label{sec3}

There are large amounts of energetic cosmic ray (CR) particles in Milk Way halo. DM can be accelerated by its interactions with those particles, obtaining velocity much larger than the escape velocity. 
We will focus on the DM-electron interaction in this work, {\it i.e.}, assuming the dark photon interaction is electron-philic
\begin{align}
\mathcal{L} \supset \epsilon \times g_{\text{em}}  A^\prime_\mu \bar{e}\gamma^\mu e ~.~ \label{eq:model2}
\end{align}
During the cosmic ray upscattering, the typical momentum transfer is much larger than the momentum of halo DM. So, we assume the halo DM to be at rest.

To give an upscattered DM with kinetic energy $T_\chi$, the minimal incoming kinetic energy of cosmic electron is~\cite{Cao:2020bwd}
\begin{equation}
T_{\mathrm{CR}}^{\min }=\left(\frac{T_{\chi}}{2}-m_{e}\right)\left[1 \pm \sqrt{1+\frac{2 T_{\chi}}{m_{\chi}} \frac{\left(m_{e}+m_{\chi}\right)^{2}}{\left(2 m_{e}-T_{\chi}\right)^{2}}}\right]~,~
\end{equation} 
where the + (-) sign is for $T_\chi>2m_e$ $(T_\chi<2m_e)$.

The recoil flux of CR-induced DM (CRDM) is obtained by convoluting the flux of cosmic electrons $d\Phi_e / dT_\mathrm{CR}$ with the DM spectrum $d \sigma_{\chi e} / d T_{\chi}$ of fixed incident electron energy~\cite{Bondarenko:2019vrb}
\begin{equation}\label{flux}
\frac{d \Phi_{\chi}}{d T_{\chi}}=D_{\mathrm{eff}} \frac{\rho_{\chi}^{\text {local }}}{m_{\chi}} \int_{T_{\mathrm{CR}}^{\min }}^{\infty} d T_{\mathrm{CR}} \frac{d \Phi_{e}}{d T_{\mathrm{CR}}} \frac{d \sigma_{\chi e}}{d T_{\chi}}~.~
\end{equation}
In Eq.~\ref{flux}, the effective distance $D_\mathrm{eff}=8.02$ kpc is obtained by integrating along the line-of-sight to 10 kpc, assuming homogeneous CR distribution and NFW DM halo profile with $\rho^\mathrm{local}_\chi=0.4~\mathrm{GeV}~ \mathrm{cm}^{-3}$.  
The differential flux of cosmic electrons $d\Phi_e / dT_\mathrm{CR} = 4 \pi dI / dT_\mathrm{CR}$, where the differential intensity of local electrons $dI / dT_\mathrm{CR}$ is simulated by HelMod-4~\cite{Boschini:2018zdv}.
Finally, the differential cross section for fixed CR kinetic energy $T_{\mathrm{CR}}$ is~\cite{Cao:2020bwd} 
\begin{equation}\label{cr-crossing}
\frac{d \sigma_{\chi e}}{d T_{\chi}}={g^{\prime 2} (\epsilon g_{\rm em})^{2}} \frac{2 m_{\chi}\left(m_{e}+T_{\mathrm{CR}}\right)^{2}-T_{\chi}\left(\left(m_{e}+m_{\chi}\right)^{2}+2 m_{\chi} T_{\mathrm{CR}}\right)+m_{\chi} T_{\chi}^{2}}{4 \pi\left(2 m_{e} T_{\mathrm{CR}}+T_{\mathrm{CR}}^{2}\right)\left(2 m_{\chi} T_{\chi}+m_{A}^{2}\right)^{2}}~.~
\end{equation}

\begin{figure}[htb]
\centering
\includegraphics[width=0.48\textwidth]{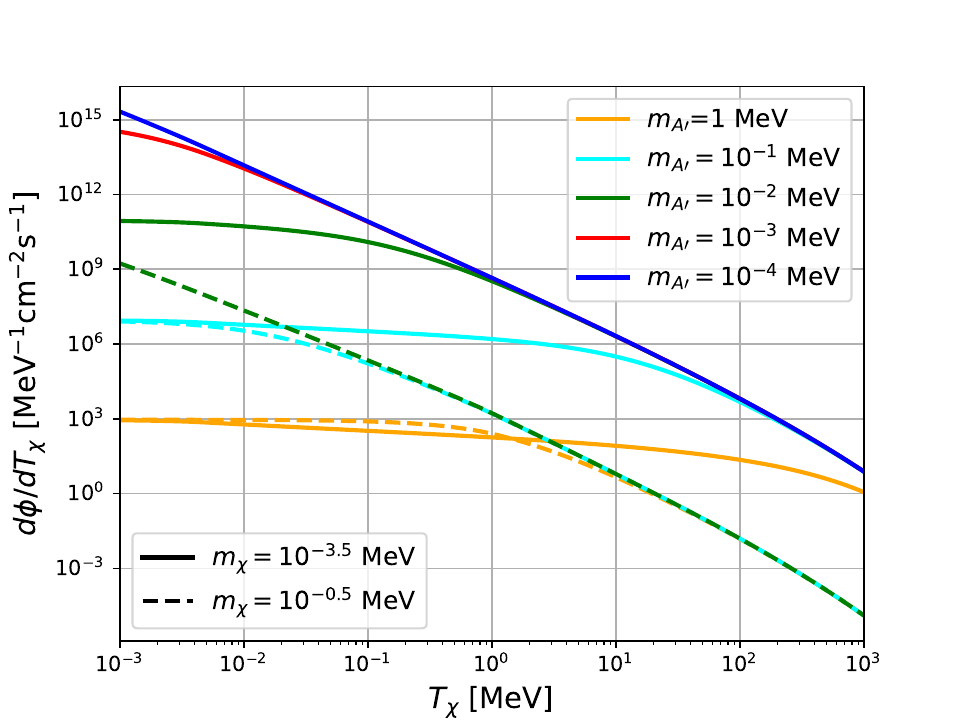}
\includegraphics[width=0.48\textwidth]{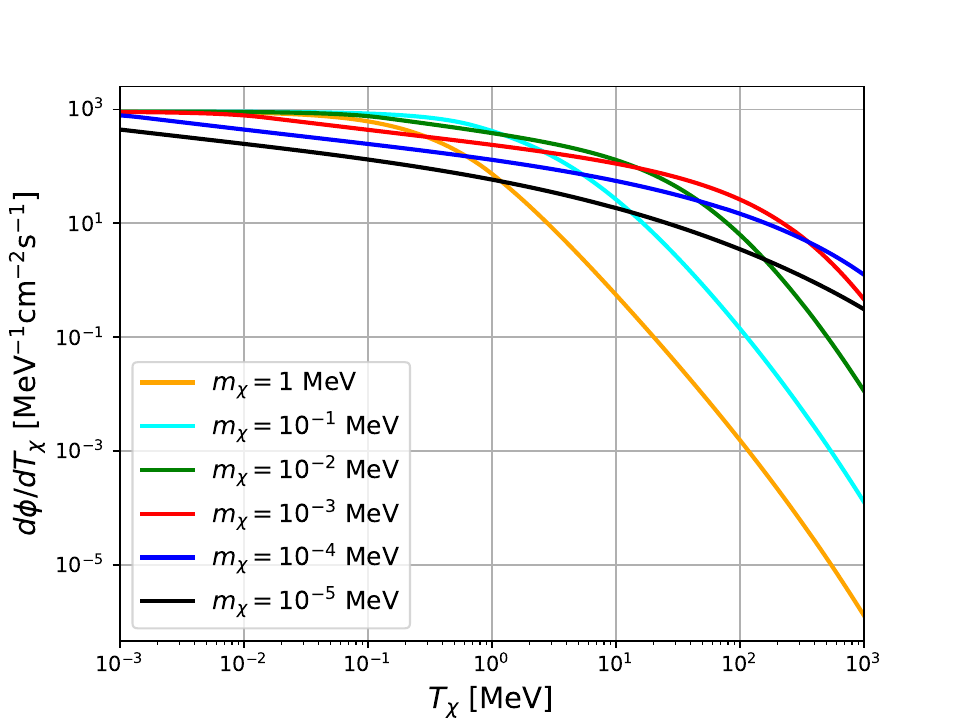}
\caption{Differential CRDM flux for different DM and dark photon masses. The masses are indicated by colors and line types as explained in the legends. In the right panel, $m_{A^\prime} = 1$ MeV is chosen. For both panels, we have chosen $g^\prime =1$ and $\epsilon=1$. 
\label{CRDM-flux}}
\end{figure}

In Figure~\ref{CRDM-flux}, we show the differential CRDM fluxes for several different choices of $m_{\chi}$ and $m_{A^\prime}$. 
In the left panel, for $m_{\chi}=10^{-3.5}$ MeV, the flux is proportional to $m_{A^\prime}^{-4}$ (for $m_{A^\prime} \gtrsim 10^{-3}$ MeV) in the low $T_\chi$ region and is independent from $m_{A^\prime}$ in the high $T_\chi$ region. 
As can be observed in Eq.~\ref{cr-crossing}, this is because the denominator (from the dark photon propagator in the matrix element) is approximate to $m_{A'}^{4}$ for small $T_\chi$ and $(2m_\chi T_\chi)^2$ for large $T_\chi$, respectively. 
Similarly, the flux becomes independent of $m_{A^\prime}$ for $m_{A^\prime} \lesssim 10^{-4}$ MeV. 
For heavier DM case, the flux becomes independent of $m_{A^\prime}$ at smaller $m_{A^\prime}$ values, as shown by the dashed curves. Note the dashed green line, dashed red line and dashed blue line are overlapping with each other. 
In the right panel, we fix $m_{A^\prime} =1$ MeV and decrease $m_\chi$ from 1 MeV to $10^{-5}$ MeV. The sizes of fluxes are similar for all $m_\chi$ when $T_\chi \lesssim 10^{-1}$ MeV, because the dependence of the differential crossing section on $m_\chi$ is cancelled by the factor of $1/m_\chi$ in Eq.~\ref{flux}. 
For larger $T_\chi \gtrsim 10^{2}$ MeV, the flux firstly increases with decreasing $m_\chi$ and then decreases for $m_{\chi} \lesssim 10^{-4}$ MeV. 
Overall, we can find that the differential flux is more flat for lighter DM and heavier dark photon, {\it i.e.,} higher fraction of high energy DM. We will focus on this region in current work. Moreover, it should be noted that the flux is proportional to $\epsilon^2$. Although we have only shown the case for $\epsilon=1$, fluxes for different $\epsilon$ can be simply obtained by overall rescaling. 

\section{Boosted dark matter scattering with electron}\label{sec4}
\subsection{Dark matter direct detection experiments}

The traditional DM direct detections aim to search for non-relativistic halo DM. Those experiments lose the sensitivity for sub-GeV DM due to the small recoil energy. 
As has been discussed above, the CR-boosted DM can have sizable rate in some parameter space. It has been studied in many literatures that the electronic recoil data at XENON1T detector is sensitive to the MeV scale CRDM.  
The scattering between the CRDM and the electrons in xenon atom is described as follows. 

A boosted CRDM scattering off an electron in target atom ($A$) is described by the ionization process $\chi+A \rightarrow \chi+A^{+}+e^{-}$. 
One can ignore the target atom and consider the initial electron as bound state within the potential of target atom. 
And the final state electron can be treated as a free particle. 
So, the process is simplified to $\chi(p_1)+e^-(p_2) \rightarrow \chi(k_1)+e^{-}(k_2)$. Those $p_{1,2}$ and $k_{1,2}$ in the brackets indicate the four-momentum of particles. 

The initial bounded electrons are in energy eigenstates, with binding energy $E_B^{nl}$ for $(n,l)$ electron shell of xenon atom. 
And their three-momentum distributions are given by momentum-space atomic wave function $\psi_{nlm}({\bf{p}})$, which can be obtained by solving Schrodinger equation with given potentials. 
So, the masses of the bounded electrons can be effectively written as~\cite{Whittingham:2021mdw} 
\begin{equation}\label{meff}
m_\mathrm{eff}^2=(m_e-E_B^{nl})^2-\bf{p}^2~.~
\end{equation} 

The differential cross section with respect to the electron recoil kinetic energy $T_R$ can be expressed by 
\begin{align}
\frac{d \sigma_{nl}}{d\ln T_R} =\frac{2l+1}{16 \cdot (2\pi)^5} & \frac{T_R |\bf{p_2}|}{E_\chi(m_e-E_B^{nl}) |{\bf p_1}| } \left|i M\left({p_{1}}, {p_{2}},{k_{1}}, {k_{2}}\right)\right|^{2} \nonumber \\
&\times  |\chi_{n l }(|{\bf p_2}|)|^2d\phi_{p_2} d| {\bf p_2}| dq ~,~ \label{sigma1}
\end{align}
where $E_\chi$ ($|{\bf p_1}|$) is incident CRDM energy (momentum size),  $\phi_{p_2}$ ($|{\bf p_2}|$) is the initial electron azimuthal angle (momentum size), and $q$ is the size of momentum transfer in the scattering. 
The $\chi_{nl}(|{\bf p_2}|)$ is the corresponding radial wave function in momentum space for electron in $(n,l)$ shell of xenon atom, whose information is available in the reference~\cite{Bunge:1993jsz}. 
The scattering amplitude $M\left({p_{1}}, {p_{2}},{k_{1}}, {k_{2}}\right)$ describes the underlying new physics. 
Note that in literature such as reference~\cite{Cao:2020bwd}, assuming initial electron to satisfy $p_2^2 = m_e^2$, only two parameters $T_\chi$ and $q$ need to be integrated. While in our case, we consider the initial electron to have effective mass different from $m_e$, so there are two additional free parameters to integrate. We take those free parameters to be $\phi_{p_2}$ and $|{\bf p_2}|$. 
More details about the calculation of the amplitude and integration ranges can be found in the Appendix~\ref{appendix-direct}

Finally, the ionization rate~\cite{Bondarenko:2019vrb} can be obtained by convoluting the CRDM flux with the differential cross section
\begin{equation}\label{recoil}
\begin{split}
\frac{dR_{ion}}{d\ln T_R}&=\sum_{nl}N_T\int dT_\chi\frac{d\sigma_{nl}}{d\ln T_R}\frac{d\phi_\chi}{dT_\chi}~,~
\end{split}
\end{equation} 
where the $N_T$ corresponds to the total number of target atoms in the detector.

\subsection{Neutrino detector - higher threshold}
Comparing to the DM direct detection experiments, the neutrino detectors probe the recoiled electron with much higher kinetic energy threshold, typically larger than $\mathcal{O}(10)$ MeV.
We will consider the probing prospects of Super-K, DUNE and JUNO experiments to the CRDM in the dark photon model of Eq.~\ref{eq:model1} and Eq.~\ref{eq:model2}.

We will mainly focus on the parameter region, where the masses of the DM and dark photon are much smaller than the typical energy scale of the DM-electron scattering in neutrino detectors. 
So, the consideration of the DM PDFs becomes necessary. 
Due to the interaction in Eq.~\ref{eq:model1}, a boosted DM can induce dark photon and anti-DM densities as well. The parton density for each dark sector species can be obtained by solving the DGLAP equation. 
All of the components will contribute to the DM-electron scattering. 
Following the factorization theorem, the differential cross section can be written as 
\begin{equation}
\frac{d \sigma}{d \ln T_{R}}=\sum_{i} \int_{0}^{x_{\max}} d x \frac{d \sigma^{i}}{d \ln T_{R}} f_{i}\left(Q, x\right) \Theta(xE^0_\chi-E_i^{\min}) ~.~
\end{equation}
The index $i$ runs over DM ($\chi$), anti-DM ($\bar{\chi}$) and dark photon ($A^{\prime}$), which correspond to the $\chi+e^-\rightarrow \chi +e^-$, $\bar{\chi}+e^-\rightarrow \bar{\chi} +e^-$ and $A^{\prime} +e^- \rightarrow \gamma +e^-$ scattering processes, respectively. 
$E_i^{\min}$ is the minimal energy of the component $i$ in incoming DM for producing an electron with recoil kinetic energy $T_R$. 
Parton density $f_i(Q,x)$ is obtained by solving DGLAP equations as discussed in the section~\ref{sec2} and the evolution scale $Q$ is defined as the sum of momentum sizes of final states, {\it e.g.,} for $\chi {(p_1)} +e^- {(p_2)}\rightarrow \chi {(k_1)}+e^- {(k_2)}$, $Q=|\vec{k}_1|+|\vec{k}_2|$. 
The lower limit of energy fraction $x$ is determined by $xE^0_\chi-E_i^{\min}>0$.
The upper limit of $x$ ($x_{\max}$) is related to the $E^0_\chi$ and masses. In the limit of $E^0_\chi \gg m_{\chi, A^{\prime}}$, it can reach $x_{\max} = 1$. 

The ionization rate of Eq.~\ref{recoil} can be generalized as 
\begin{align}
\frac{dR_{ion}}{d\ln T_R}=&N^{SK}_T\sum_i \int dT^0_\chi\int_0^{x_{\max}}dx\frac{d\sigma^i}{d\ln T_R}f_i(Q,x)\frac{d\phi_\chi}{dT^0_\chi}\Theta(xE^0_\chi-E_i^{\min}) \nonumber
\\
=&N^{SK}_T\sum_{i} \int dT_i\int^{x_{\max}}_{0}dx\frac{1}{x}\frac{d\sigma^i}{d\ln T_R}f_i(Q,x)\frac{d\phi_\chi}{dT^0_\chi}\Theta(T_i+m_i-E_i^{\min}) ~,~\label{sk-recoil}
\end{align}
where a new variable $T_i=xE_\chi^0-m_i$ which is the kinetic energy of component $i$ in incoming DM has been introduced in the second line. 
Moreover, the upper limit of the energy fraction $x$ for $\chi$ ,$\bar{\chi}$ and $A^{\prime}$ can be calculated by 
\begin{equation}\label{xmax}
\begin{split}
&\frac{T_\chi+m_\chi}{x}(1-x)>m_{A^\prime}~,~
\\
&\frac{T_{\bar{\chi}}+m_\chi}{x}(1-x)>2m_\chi~,~
\\
&\frac{T_{A^{\prime}}+m_{A^{\prime}}}{x}(1-x)>m_\chi~,~
\end{split}
\end{equation} 
respectively. Those conditions correspond to the minimal requirements for considering mass effects. 

We need to emphasize that the mass effects are ignored in the integration bounds of the DGLAP equations, in contrast to the calculation of ionization rate above. 
Since we focus on the case where the recoil energy is much larger than particle masses, such treatment is still acceptable except for the DM-electron scattering. 
Applying the first condition of Eq.~\ref{xmax} clearly removes the phase space in which the DM carries all of the initial energy, {\it i.e.,} $x_{\chi} =1$. 
To include this contribution consistently in our calculation, the integrated parton density $\int^{1}_{x_{\max}} f_\chi(Q,x) dx$ is taken as the probability of DM with $x=1$. 
As a result, the ionization rate receives an extra contribution
\begin{equation}
N^{SK}_T\int dT_\chi \frac{d\sigma_\chi}{d\ln T_R}\frac{d\phi_\chi}{dT_\chi }\Theta(T_\chi-T_\chi^{\min})\int^1_{x_{\max}}f_\chi(Q,x)~.~
\end{equation}

Considering the DM PDFs, it will be most interesting to observe the scattering between the dark photon and the electron, {\it i.e.} $A^{\prime} + e^-\rightarrow \gamma +e^-$. 
This process is also dubbed as dark Compton scattering~\cite{Hochberg:2021zrf}, in which one can observe the photon in the final states besides the recoiled electron.
The corresponding recoil rate is given by
\begin{align}
\frac{dR}{d\ln E_\gamma} = &N^{SK}_T \int dT^0_\chi\int_0^{x_{\max}}dx\frac{d\sigma^{A^{\prime}}}{d\ln E_\gamma}f_{A^{\prime}}(Q,x)\frac{d\phi_\chi}{dT^0_\chi}\Theta(xE^0_\chi-E_{A^{\prime}\gamma}^{\min})\Theta(E_{A^{\prime}\gamma}^{\max}-xE^0_\chi) \nonumber
\\
= & N^{SK}_T \int dT_{A^{\prime}}\int^{x_{\max}}_{0}dx\frac{1}{x}\frac{d\sigma^{A^{\prime}}}{d\ln E_\gamma}f_{A^{\prime}}(Q,x)\frac{d\phi_\chi}{dT^0_\chi} \nonumber \\
& \times \Theta(T_{A^{\prime}}+m_{A^{\prime}}-E_{A^{\prime}\gamma}^{\min}) \cdot \Theta(E_{A^{\prime}\gamma}^{\max}-T_{A^{\prime}}-m_{A^{\prime}})~,~
\end{align}
where $E_{A^{\prime}\gamma}^{\min}$ and $E_{A^{\prime}\gamma}^{\max}$ are given in Eq.~\ref{range-sk-3}, and $x_{\max}$ is constrained by the third equation in Eqs.~\ref{xmax}.
More details of the calculation are provided in the Appendix~\ref{appendix-sk}.

\section{Results}\label{sec5}
\subsection{Bounds from XENON1T and SuperK}

We have calculated the recoil rate of DM-electron scattering at both DM direct detection experiments and neutrino detectors, in terms of parameters $m_{\chi}$, $m_{A^{\prime}}$, $g^{\prime}$ and $\epsilon$. 
In this section, considering the measurements of electronic recoil data at the XENON1T detector and the Super-Kamiokande as examples, we calculate the corresponding bounds for our simplified model. 

For XENON1T experiment, the consistency between the theory and experiment is tested by the $\chi^{2}$~\cite{Li:2022jxo}, which is defined as
\begin{equation}
\chi^{2}=\sum_{i} \frac{\left[\left(\frac{d R_{\chi+B_{0}}}{d T_{R}}\right)_{i}-\left(\frac{d R_{o b s}}{d T_{R}}\right)_{i}\right]^{2}}{\sigma_{i}^{2}}~,~
\end{equation}
where $\frac{d R_{\chi}}{d T_{R}}$,  $\frac{d R_{B_0}}{d T_{R}}$ and $\frac{d R_{o b s}}{d T_{R}} $ are the recoil rates in the $i$th bin from theoretical prediction, estimated background and the observation, respectively. 
The $\sigma_i$ in the denominator corresponds to the uncertainty in the $i$th bin.
The event numbers of background and observation in each bin can be found in reference~\cite{XENON:2020rca}. 
The $\chi^{2}$ value for background only is 46.4. Assuming the test statistic follows a $\chi^{2}$ distribution with two degrees of freedom, the $2\sigma$ bound corresponds to $\Delta \chi^2=\chi^2-\chi_{B_0}^2=6.18$.

For Super-K experiment, an analysis for boosted DM search with electron recoil kinetic energy $T_{e} >100$ MeV has been performed on the data corresponding to 161.9 kiloton-year exposure~\cite{Super-Kamiokande:2017dch}.
The total measured number of events $N_{\rm sk}$ is 4042 in the bin $0.1<T_e/\text{GeV}<1.33$. 
Following the analysis proposed in reference~\cite{Ema:2018bih}, a conservative upper limit on DM signal can be obtained by requiring 
\begin{equation}\label{sk-exclusion}
\xi \times N_{\rm DM}<N_{\rm sk}~,~
\end{equation}
where the signal efficiency $\xi=0.93$. 
The number of DM scattering events $N_{\rm DM}$ is calculated by integrating $\frac{dR_{ion}}{dT_R}$ over the region $T_R>100~\mathrm{MeV}$, with total number of electrons inside the Super-K detector $N_{e}=7.5\times10^{33}$ and data-taking period of 2628.1 days. 

\begin{figure}[htb]
\centering
\includegraphics[width=0.48\textwidth]{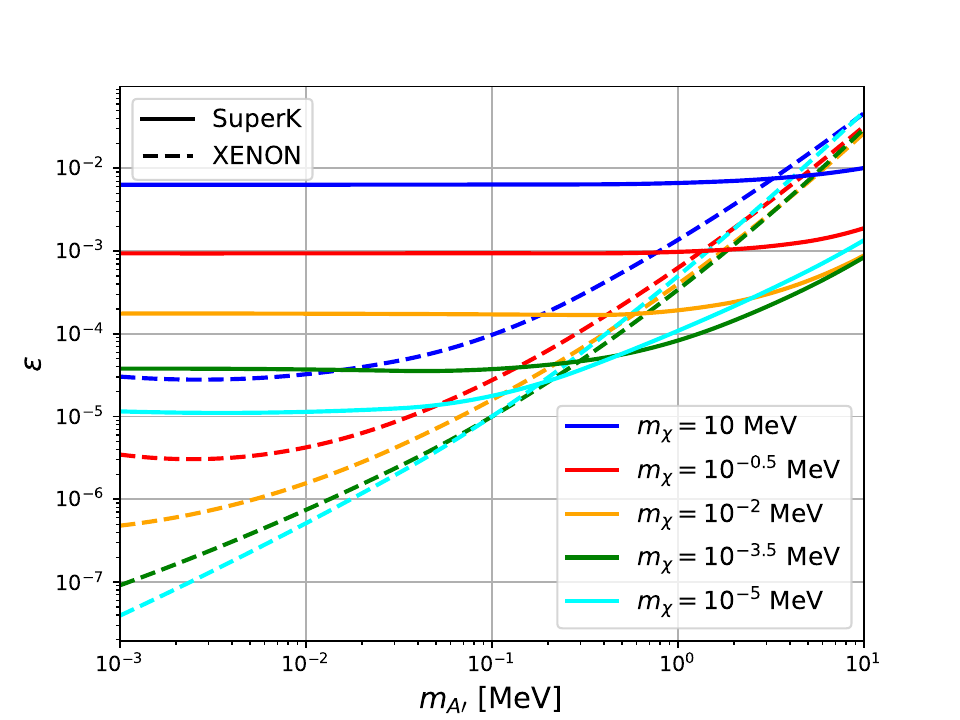}
\includegraphics[width=0.48\textwidth]{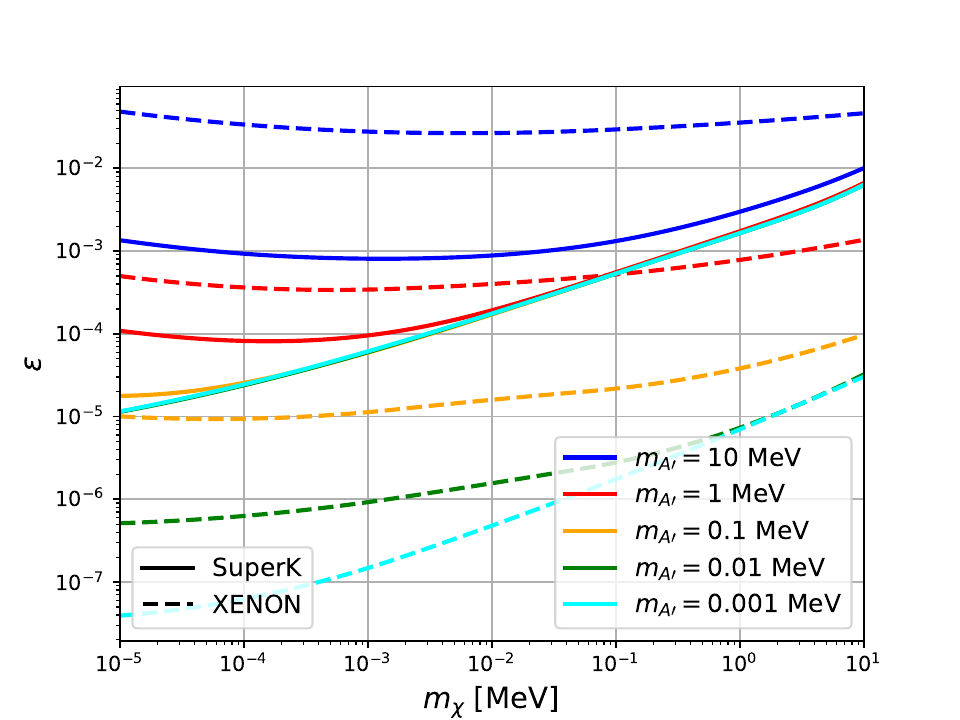}
\caption{Left panel: exclusion limits of XENON1T (dashed lines) and Super-K (solid lines) in the $m_{A^{\prime}}-\epsilon$ plane. The values of $m_{\chi}$ are indicated by the line colors. Right panel: exclusion limits of XENON1T (dashed lines) and Super-K (solid lines) in the $m_{\chi}-\epsilon$ plane. The values of $m_{A^{\prime}}$ are indicated by the line colors. In both cases, the DM coupling $g^{\prime}=1$. \label{exclusion-limit}}
\end{figure}

In Figure~\ref{exclusion-limit}, we plot the exclusion limits for XENON1T and Super-K in the $m_{A^{\prime}}-\epsilon$ plane and $m_{\chi}-\epsilon$ plane. 
In the left panel, it can be observed that the sensitivities of both experiments degrade with increasing dark photon mass, and the degradation of XENON1T sensitivity is severer.  
This feature can be attributed to the dark photon propagators for both CR-DM scattering and DM-electron scattering. 
The dark photon propagator contains terms of $m_{A'}^2$, $2m_\chi T_\chi$ and $2m_e T_R$. 
For the recoil rate, each of the two propagators will be suppressed by $1/ m_{A'}^{2}$ when $m_{A'}$ dominates over other terms. 
The typical energy scale of DM-electron scattering at Super-K experiment is much larger than that at XENON1T experiment. So, the $m_{A'}$ suppression for Super-K occurs at larger $m_{A'}$. 
In the right panel, given the dark photon mass, both experiments are most sensitive to a certain $m_{\chi}$ (the specific value depends on $m_{A^{\prime}}$), while the sensitivities degrade for  $m_{\chi}$ deviating from this value. 
This feature is mainly attributed to the flux of CRDM, as we have been discussed for the right panel of Figure~\ref{CRDM-flux}, {\it i.e.,} for a fixed $m_{A^{\prime}}$, in the kinematic region $T_{\chi} \gtrsim \mathcal{O}(1)$ MeV, the CRDM flux is highest for a certain value of $m_{\chi}$. 

Moreover, in the Figure~\ref{exclusion-limit}, we can find that the Super-K experiment is complementary to the XENON1T experiment in exploring the full parameter space.
And the Super-K experiment exhibits a stronger bound than the XENON1T experiment in the large $m_{A^{\prime}}$ and light $m_{\chi}$ region. 
From the left panel, the intersection points (between the solid and dashed lines with the same color) show positive correlation between $m_\chi$ and $m_{A'}$.
As has been discussed before, for the CRDM flux, the fraction of energetic DM is higher for heavier dark photon. 
Thus, in the heavy dark photon region, the signal rate at Super-K detector is less suppressed by the $m_{A^{\prime}}$ than that at XENON1T. 
Nevertheless, the situation becomes opposite in the light dark photon region, where most CRDM have very small energy, rendering better sensitivity of XENON1T detector.

\begin{figure}[htb]
\centering
\includegraphics[width=0.7\textwidth]{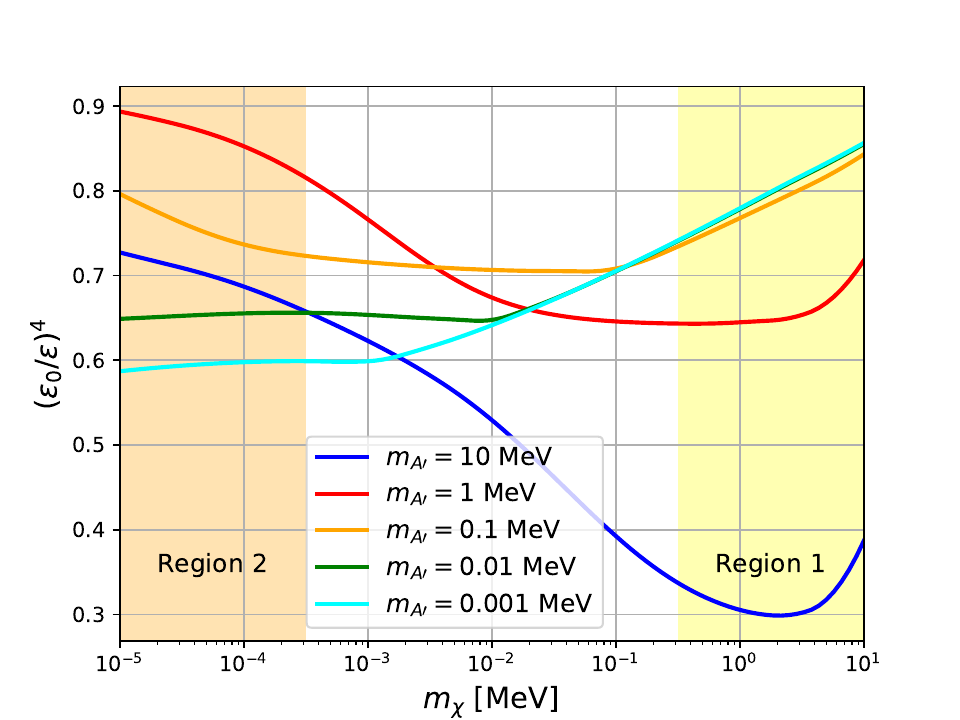}
\caption{The ratio between the exclusion limits of Super-K with (denoted by $\epsilon$) and without (denoted by $\epsilon_0$) considering the PDF effects. 
The ratios for different dark photon masses are indicated by the line colors. 
The dark matter coupling $g^{\prime}=1$. 
\label{ISR-SK}}
\end{figure}

\begin{figure}[htb]
\centering
\includegraphics[width=0.45\textwidth]{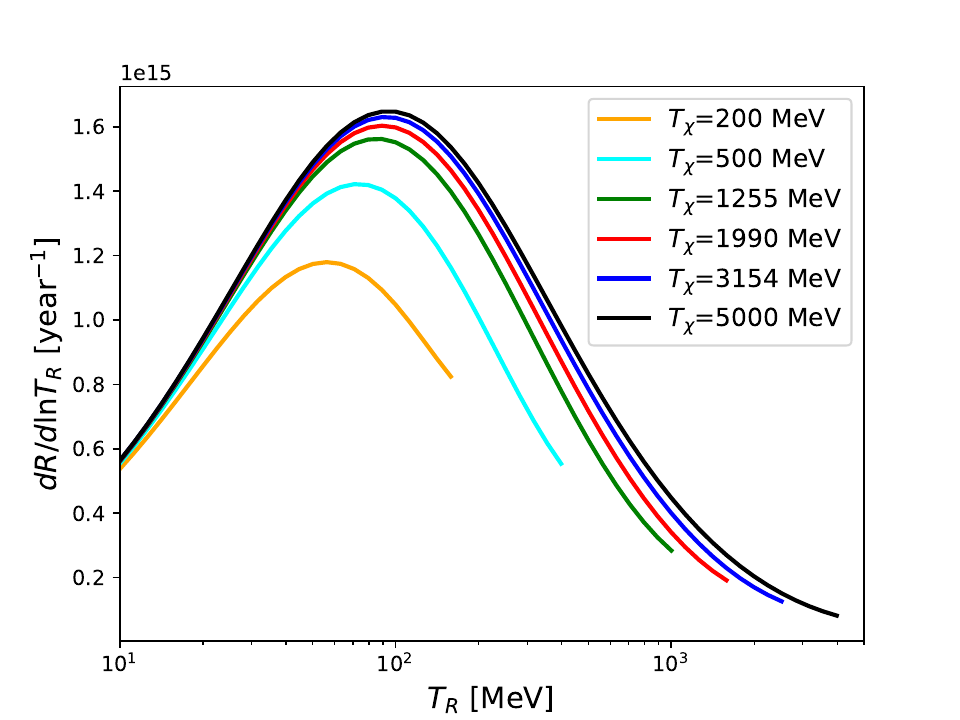}
\includegraphics[width=0.45\textwidth]{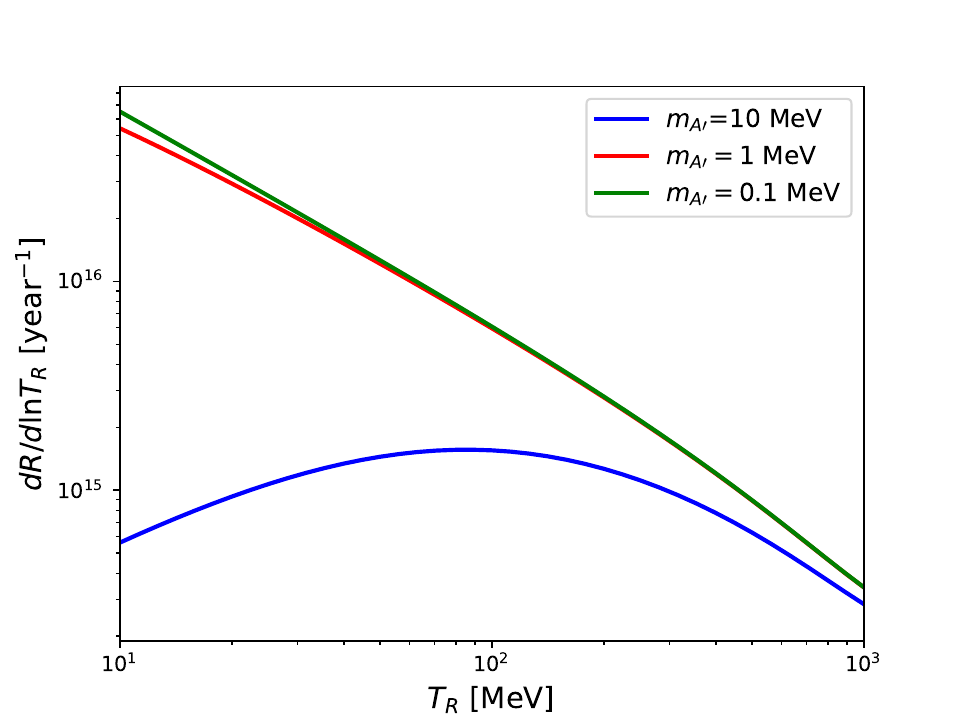}
\caption{Left panel: the differential recoil rate with fixed incident DM kinetic energy as indicated in the legend, where we have chosen $m_\chi=0.32$ MeV, $m_{A^{\prime}}=10$ MeV, $g^{\prime}=1$ and $\epsilon=1$. Right panel: the differential recoil rate for three different values of $m_{A^{\prime}}$, where we have taken incident DM kinetic energy $T_\chi=1255$ MeV, $m_\chi=0.32$ MeV,  $g^{\prime}=1$ and $\epsilon=1$. \label{characteristic}}
\end{figure}

To emphasize the importance of the DM PDF effects for DM detection at neutrino detector, we plot the ratio between the exclusion limits of Super-K with (denoted by $\epsilon$) and without (denoted by $\epsilon_0$) considering the PDF effects in Figure~\ref{ISR-SK}. 
The ratio is shown by $(\epsilon_0 / \epsilon )^4$ since the recoil rate 
$\frac{dR_{ion}}{d\ln T_R}  \propto \epsilon^{4}$. 
The recoil rate without DM PDF effects is calculated from Eq.~\ref{sk-recoil}, by taking into account only the DM component and setting parton density $f_\chi=\delta(1-x)$ and $f_{A^{\prime}}=f_{\bar{\chi}}=0$.
In the full parameter space, the DM PDF effects always reduce the experimental sensitivity, {\it i.e.,} the allowed value of $\epsilon$ is always larger than that of $\epsilon_{0}$ for given DM and dark photon masses. 
The reason is twofold.   
Firstly, for the DM-electron scattering, the recoil rate is higher for larger incident DM energy. To demonstrate this point, in the left panel of Figure~\ref{characteristic}, we plot the differential recoil rate for several different incident DM kinetic energies, with fixed $m_{\chi}$, $m_{A^{\prime}}$, $g^{\prime}$ and $\epsilon$. The features are kept the same for other parameter choices. 
Including the PDF effects will soften the $\chi$ spectrum, thus leading to the reduced recoil rate. 
Secondly, due to the relatively soft CRDM flux and high threshold of Super-K ($T_{R} > 100$ MeV), the dark photon and anti-DM components can only provide sub-dominant contributions to the electron recoil. 
As has been discussed above, the CRDM flux deceases quickly for $T_{\chi} \gtrsim 100$ MeV in the parameter space of interest and the dark photon/anti-DM density is sizable only in the small $x$ region. 
  
The feature of ratios in Figure~\ref{ISR-SK} can be understood better in terms of two mass limits. The marked region 1 and region 2 in the figure correspond to heavy DM region and light DM region, respectively. 
In the region 1, the CRDM flux (in the high $T_{\chi}$ region) is almost the same for different $m_{A^{\prime}}$, as shown by dashed lines in the left panel of Figure~\ref{CRDM-flux}. 
On the contrary, as shown in the right panel of Figure~\ref{characteristic}, the fraction of events with $T_{R} >100$ MeV is higher for heavier dark photon, especially when $m_{A^{\prime}} \gtrsim 1$ MeV.  
This means that the typical energy scale of DM-electron scattering for heavier dark photon case is higher, leading to more significant PDF effects. 
As the dark photon/anti-DM contributions are subdominant and the energy spectrum of DM component is softened by the PDF effects, the recoil rate is strongly suppressed when the dark photon is heavy. 
In the region 2, the ratio of recoil rates is most close to 1 when $m_{A^{\prime}} \sim 1$ MeV and decreases as the $m_{A^{\prime}}$ deviates from 1 MeV. 
For dark photon mass heavier than 1 MeV, the ratio decreases with increasing $m_{A^{\prime}}$, due to the similar reason as discussed for region 1. 
As for dark photon mass below 1 MeV, the feature is mainly determined by the flux of CRDM. 
The CRDM flux is more flat for heavier $m_{A^{\prime}}$ as shown in the left panel of Figure~\ref{CRDM-flux}, {\it i.e.,} higher fraction of energetic DM. 
The distribution of recoil energy has strong dependence on the DM kinetic energy in the region 
$T_\chi \lesssim 10^3$ MeV, as shown in the left panel of Figure~\ref{characteristic}.
So, the PDF effects are most significant when the signal DM flux is dominated by the DM with $T_\chi \sim \mathcal{O}(100)$ MeV (this is the case for lighter dark photon).   
Furthermore, the flux also becomes more flat when $m_\chi$ decreases, so the ratio increases as decreasing $m_\chi$.

\subsection{The photon signal at neutrino detectors}

Besides the recoiled electron signal, the dark photon component can induce mono-energetic photon final state through the dark Compton scattering $A^{\prime} e \to \gamma e$. 
It has been found that the dark Compton scattering has a substantial impact on the reach and discovery potential of direct detection experiments for bosonic DM~\cite{Hochberg:2021zrf}.

\begin{figure}[htb]
\centering
\includegraphics[width=0.48\textwidth]{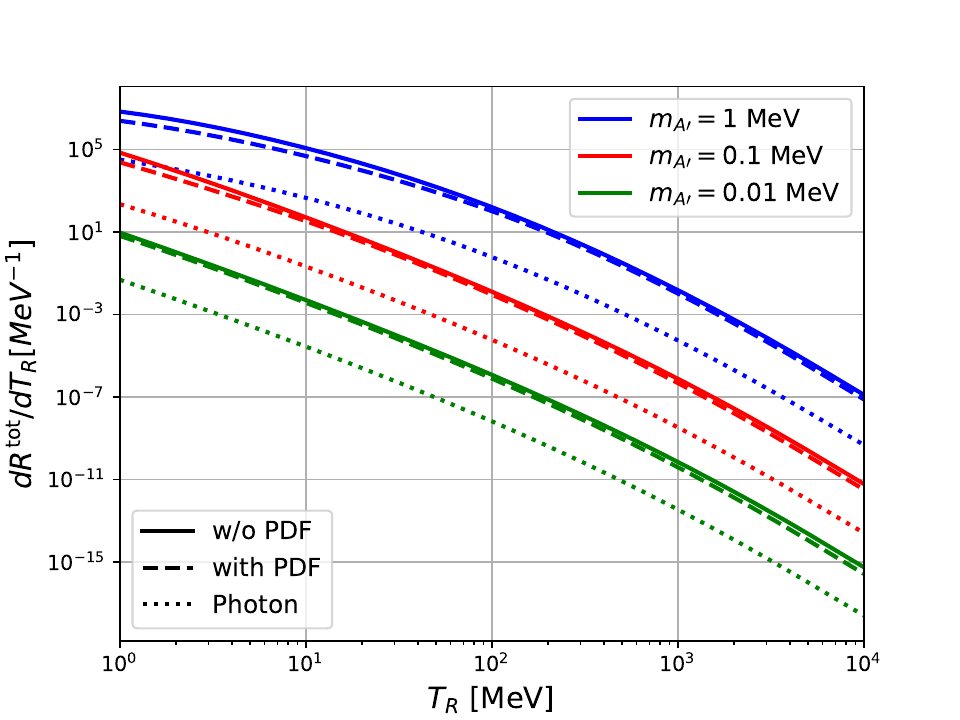}
\includegraphics[width=0.48\textwidth]{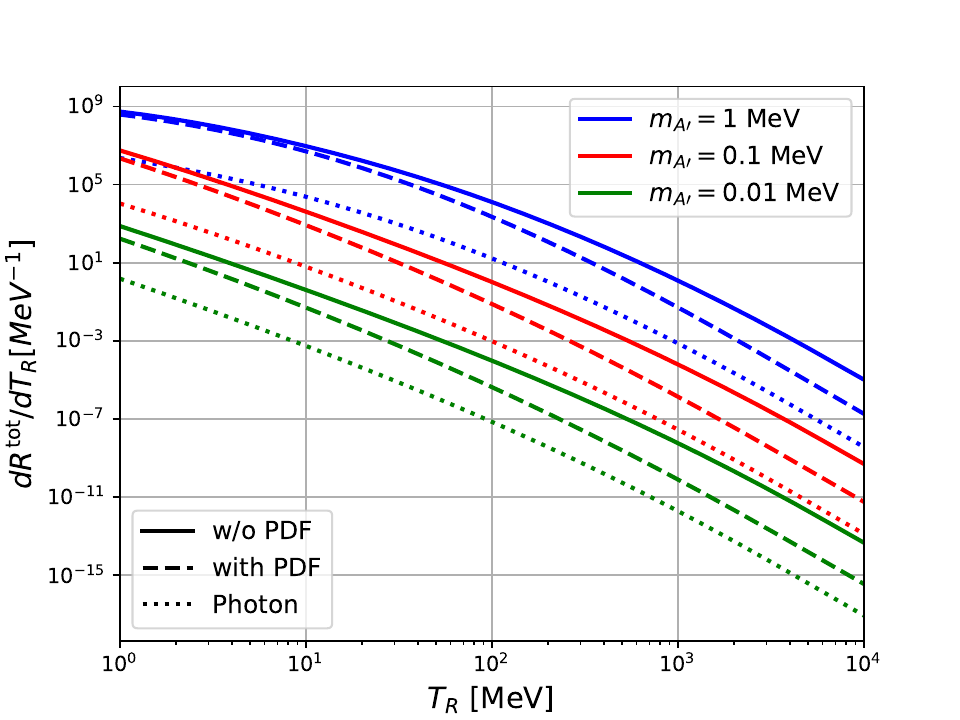}
\caption{The differential recoil rate (at Super-K with data taking of 2628.1 days) for recoiled electron without DM PDF effects (solid line), with DM PDF effects (dashed line) and for the outgoing photon (dotted line), where we have taken fixed dark photon mass as indicated in the legend, DM mass $m_\chi=0.01$ MeV, signal efficiency $\xi=0.93$ and the $\epsilon$ is chosen as the maximal value that satisfies the XENON1T and Super-K bounds. Left panel: DM coupling $g^{\prime}=1$; Right panel: DM coupling $g^{\prime}=3$. 
 \label{signal}}
\end{figure}

In Figure~\ref{signal}, we show the differential recoil rate for the outgoing photon induced by the dark photon component in the DM PDF at the Super-K detector, where we also plot the corresponding differential recoil rate for the recoiled electron for comparison. 
The parameter choices are explained in the caption, in particular the $\epsilon$ values are chosen as the maximal values that satisfy the XENON1T and Super-K bounds, which are $1.92\times10^{-4}$, $1.59\times10^{-5}$ and $1.56\times10^{-6}$ for $m_{A^{\prime}}=1$ MeV, $m_{A^{\prime}}=0.1$ MeV and $m_{A^{\prime}}=0.01$ MeV, respectively. 
In the left panel, the DM coupling is chosen as $g^{\prime}=1$ such that the DM PDF effects are mild. 
For the $m_{A^{\prime}}=1$ MeV case (blue lines), the Super-K poses a stronger bound than the XENON1T due to the electron recoil. 
At the Super-K, it can produce 4042 recoiled electrons with $T_{R} > 100$ MeV and 40519.8 outgoing photons with $T_{R} > 1$ MeV. 
For the $m_{A^{\prime}}=0.1$ MeV case (red lines), the XENON1T poses a stronger bound instead. At the Super-K, such parameter choice can lead to 0.30 recoiled electrons with $T_{R} > 100$ MeV and 119.56 outgoing photons with $T_{R} > 1$ MeV.
As for the third case $m_{A^{\prime}}=0.01$ MeV, XENON1T constraint is much stronger than the Super-K constraint. The numbers of both recoiled electrons and outgoing photons are small at Super-K detector, {\it i.e.,} $2.56\times 10^{-5}$ and 0.022 for electrons with $T_{R} > 100$ MeV and photons with $T_{R} > 1$ MeV. 
The rate of mono-energetic photon can be further enhanced for larger DM coupling $g^{\prime}$ as shown in the right panel of Figure~\ref{signal}~\footnote{For comparison purpose, the $\epsilon$ values are kept the same as those for $g^{\prime}=1$.}. 
Firstly, the flux of CRDM is proportional to $g^{\prime 2}$. The rates of both the signal electron and photon are enhanced, leading to a stronger limit on $\epsilon$. 
Secondly, although the ratio $(\epsilon_{0} / \epsilon)^{4}$ (as shown in Figure~\ref{ISR-SK}) is always greater than $\mathcal{O}(0.1)$ for $g^{\prime}=1$, it can be reduced to $\lesssim \mathcal{O}(10^{-2})$ for $g^{\prime}=3$. 
It means that the calculated exclusion limits without considering the DM PDFs are considerably over-estimated. 
Thirdly, the ratio between the numbers of signal photons and signal electrons can be significantly enhanced by the large $g^{\prime}$. 
As a result, the mono-energetic photon signal may provide a much better experimental sensitivity than the recoiled electron when $g^{\prime}$ is large.

However, the Super-K is water-based Cherenkov detectors, where the Cherenkov rings induced by photons and electrons are similar. It is not possible to identify the mono-energetic photon with threshold $\mathcal{O}(1)\sim\mathcal{O}(10)$ MeV. 
On the other hand, as discussed in reference~\cite{Cui:2022owf}, DUNE and JUNO detectors have the capacity of high efficiency photon identification, which enable us to identify the mono-energetic photon with a much lower threshold. 
Moreover, such an energetic single photon signal at DUNE and JUNO can be treated as background free, which means an event rate of a few can lead to the exclusion/discovery. 

\begin{figure}[htb]
\centering
\includegraphics[width=0.48\textwidth]{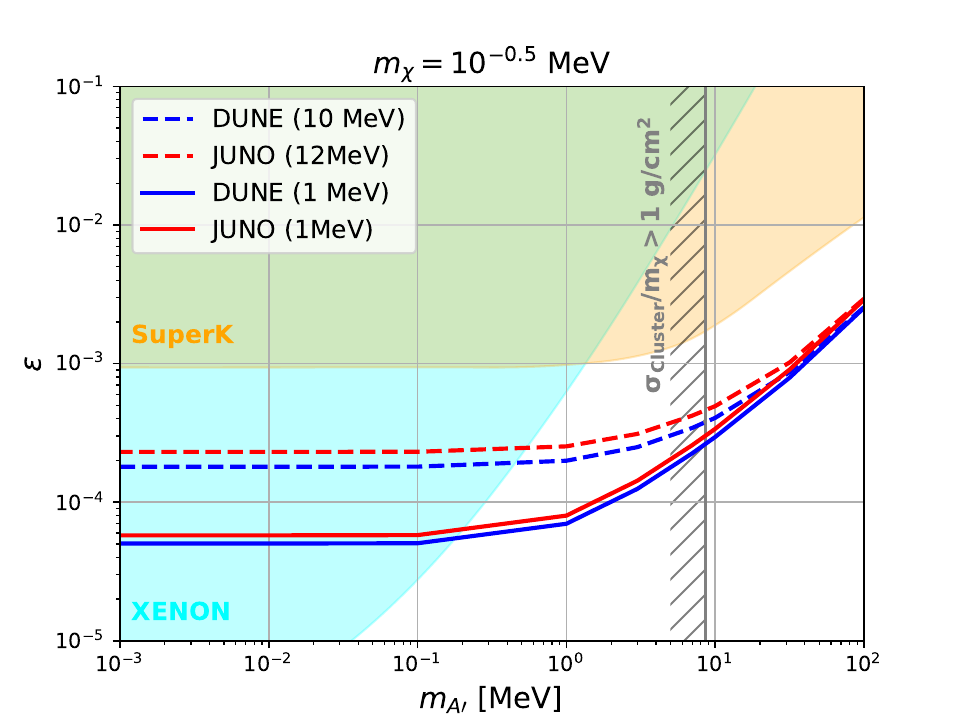} 
\includegraphics[width=0.48\textwidth]{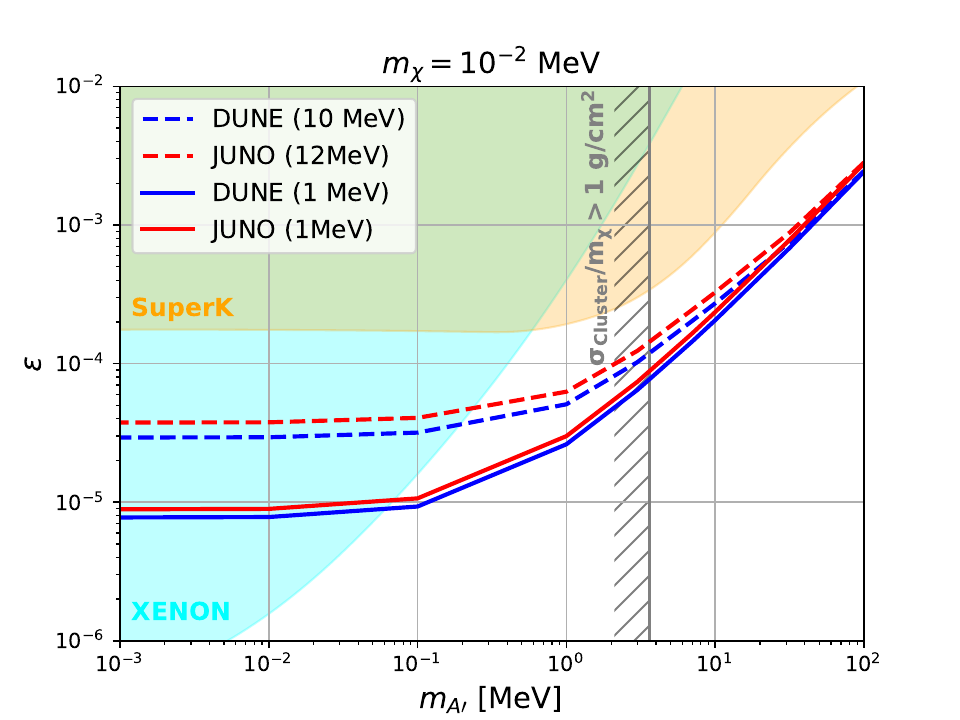}\\
\includegraphics[width=0.48\textwidth]{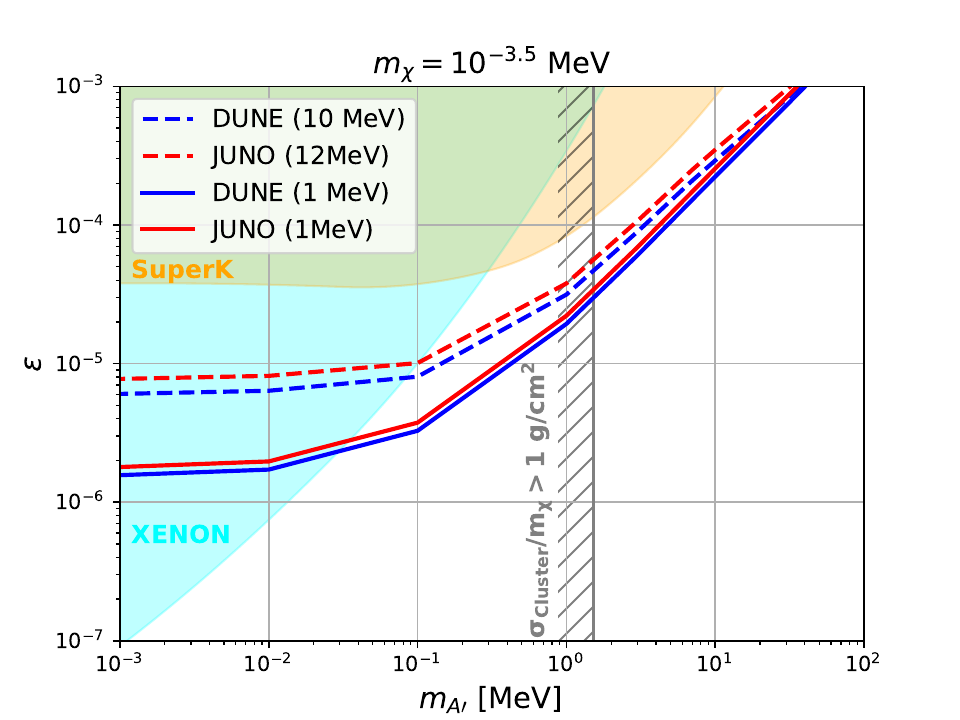}
\includegraphics[width=0.48\textwidth]{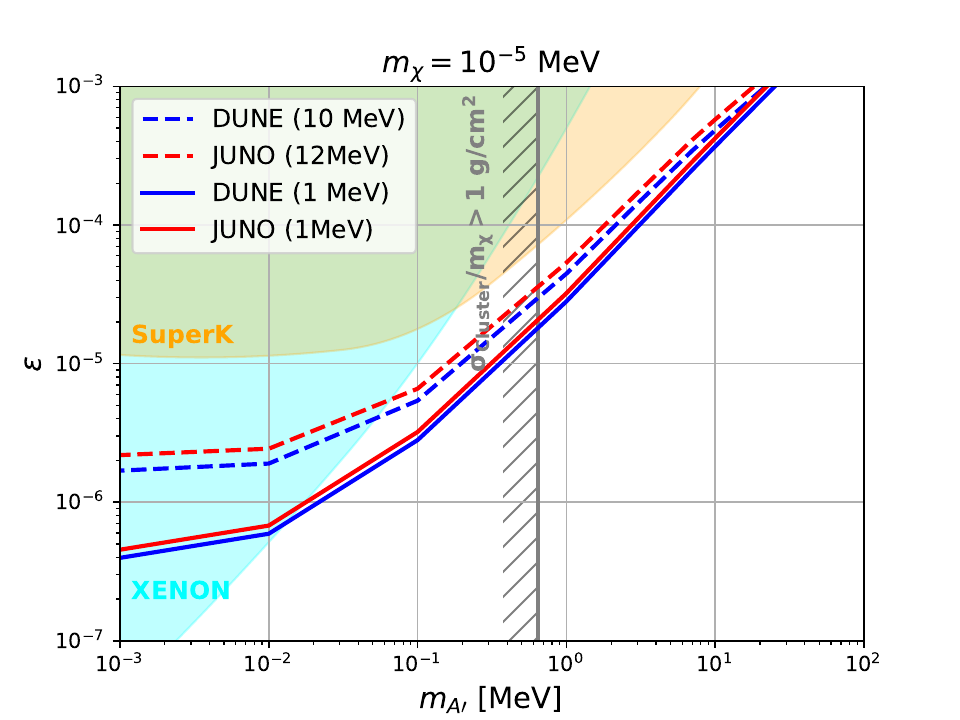}
\caption{The orange shaded regions and cyan shaded regions correspond to the Super-K and XENON1T bounds. 
The DM self-interaction constraints are shown by hatched vertical lines, and regions on the left-hand side are excluded.
The blue lines and red lines correspond to the contours of three signal photon events in each year at DUNE and JUNO, respectively.
The threshold of the photon detection is indicated by the line types as explained in the legend. We choose four different DM masses for demonstration, the values of which are given in the titles of plots. \label{DUNEJUNO}
 }
\end{figure}

In Figure~\ref{DUNEJUNO}, we plot the contours of three signal photon events per year at DUNE and JUNO experiments in the $m_{A^{\prime}}-\epsilon$ plane, assuming two different photon energy thresholds for each experiment. 
For DUNE detector, considering the single phase LArTPC module, we calculate the sensitivity reach with active LAr of 40 kilotons, which corresponds to $1.085\times 10^{34}$ electrons inside the detector. 
The JUNO experiment will be equipped with liquid scintillator (2,5-diphenyloxazole) detector with fiducial mass of 20 kilotons, so that the total number of electrons inside the detector is $6.314\times 10^{33}$. 
For simplicity, we assume the detection efficiencies of energetic photon of both detectors to be unity. 
From Figure~\ref{DUNEJUNO}, we can find that the features of DUNE and JUNO sensitivities are similar, and the DUNE provides slightly better sensitivity than the JUNO due to a larger number of electrons in the detector. 
Figure~\ref{DUNEJUNO} also shows that the photon component in DM PDFs provides an efficient and complementary way to discover the DM sector, especially in the parameter space with relatively heavy dark photon and light DM. 
It should be noted that although we have shown the exclusion curves of DUNE and JUNO with dark photon mass up to 100 MeV, the limits in the region $m_{A^{\prime}} \gtrsim \mathcal{O}(10)$ MeV may suffer from large uncertainty. Because the condition of $E_{\chi} \gg m_{A^{\prime}}$ in PDF calculation (see discussions in section~\ref{sec2}) may not be fully satisfied. 

\subsection{Constraints on DM self-scattering cross section}

{
The light mediator as well as sizeable coupling lead to remarkable DM self-scattering, the cross-section of which could be constrained by Bullet Cluster~\cite{Markevitch:2003at,Clowe:2003tk,Randall:2008ppe} and other cosmological simulations with self-interacting DM on the scales of galaxies and galaxy clusters~\cite{Rocha:2012jg,Peter:2012jh}~\footnote{We will not attempt to explain the ``small scale problems'' with our DM scenario, since they may be resolved within complete simulations including baryons~\cite{Duffy:2010hf,Pontzen:2011ty,Kim:2017iwr,Tulin:2017ara}.}. The general bound is $\sigma_{\text{self}} /m_{\chi} <1$ cm$^{2}/$g.
In our case, the DM self-interaction is mediated by the dark photon. The cross section of the scattering between $\chi$ and $\bar{\chi}$ in the non-relativistic limit is determined by the attractive Yukawa potential
\begin{equation}
V(r)=-\alpha^\prime \frac{e^{-m_{A^\prime} r}}{r}~,
\end{equation}
where $\alpha^\prime=g^{\prime 2}/(4\pi)$. With $\delta_l$ being the phase shift for a partial wave $l$, the scattering amplitude can be expressed as \begin{equation}
f(\theta)=\frac{2}{m_\chi v} \sum_{l=0}^{\infty}(2 l+1) e^{i \delta_l} P_l(\cos \theta) \sin \delta_l~,
\end{equation}  
where $v$ ($v\ll 1$) is the relative velocity between $\chi$ and $\bar{\chi}$ in the scattering. We use the transfer cross section $\sigma_{\text{T}}$ to describe this scattering~\cite{PhysRevA.60.2118}, which is defined as
\begin{align}
\sigma_{\mathrm{T}}&=\int d \Omega(1-\cos \theta) \left| f(\theta)\right|^2\\
&=\frac{16 \pi}{m_\chi^2 v^2} \sum_{\ell=0}^{\infty}(\ell+1) \sin ^2\left(\delta_{\ell+1}-\delta_{\ell}\right)~.
\end{align}
The characteristic velocity on cluster scales is around $1000$ km/s.
Instead of numerically solving the Schr\"{o}dinger equation with the potential $V(r)$ to obtain $\delta_l$ and then $\sigma_{\text{T}}$ with too much time cost, the improved analytical expression for the DM
self-interacting cross section which takes into account the Born level effects proposed in ~\cite{Colquhoun:2020adl,Li:2021bka} can be used to estimate $\sigma_{\text{T}}$ quickly. 

In Figure~\ref{DUNEJUNO}, the lower limits on the dark photon mass from DM self-scattering constraints are indicated by the hatched vertical lines. 
Although we have calculated the bounds with the full analytical expression, the parameter regions around the hatched lines are in the Born regime since $\alpha^\prime m_\chi/ m_{A^\prime}\lesssim 1$. In the Born regime the analytical expression for $\sigma_{\text{T}}$ can be simplified to
\begin{align}
\sigma_{\mathrm{T}}^{\text {Born }}&=\frac{8\pi \alpha^{\prime 2}}{m_\chi^2 v^4}\left(\ln \left(1+4 t^2\right)-\frac{4 t^2}{1+4 t^2}\right)\\
&\overset{t\ll 1}{=}\frac{g^{\prime 4} }{4\pi }\frac{ m_\chi^2}{m_{A^\prime}^4}~,
\end{align}
where $t=m_\chi v/(2m_{A^\prime})$. Given the dark matter coupling $g^{\prime}=1$, the lower bounds on the dark photon mass vary as $(m_{\chi})^{1/4}$, which are 8.582 MeV, 3.619 MeV, 1.526 MeV and 0.644 MeV for $m_{\chi}=10^{-0.5}~\text{MeV}, 10^{-2}~\text{MeV}, 10^{-3.5}~\text{MeV}$ and $10^{-5}$ MeV, respectively.  
}

\section{Conclusion}\label{sec6}

In this work, we study the constraints and detection prospects of the boosted DM at DM direct detection experiments and neutrino detectors. 
A simplified electron-philic dark photon model with fermionic DM is considered. 
In the model, the DM can be accelerated through its scattering with high energy CR particles. 
We find that the spectrum of CRDM flux is more flat in the region with relatively heavy dark photon and light DM, such that the DM with energy well above the detector threshold could still provide non-negligible contribution to the DM-electron scattering. 
Since the energy threshold of neutrino detectors is around $\mathcal{O}(10-100)$ MeV, the parameter space with $m_{A^{\prime}}, ~m_{\chi} \in  [1~\text{keV}, 1~\text{MeV}]$ is investigated in detail.  

In calculating the electron recoil rate at the XENON1T, we consider carefully the bound state effects of the initial electron by introducing the effective mass and atomic wave function. 
Due to the low energy threshold of the detector, it is most sensitive to the parameter space with light $m_{A^{\prime}}$ where the coupling $\epsilon$ as small as $10^{-7}$ is excluded. 

The calculation at neutrino detector is more complicated, due to the hierarchy between the mass of DM and the energy scale of the DM-electron scattering. 
It will lead to large logarithms which significantly affect the differential cross section of the DM scattering.
The DGLAP equation is adopted for resummation of those large logarithms, producing the DM PDFs. 
The high energy electron recoil signals can be induced by all of the DM, anti-DM and dark photon components in the DM PDFs, which are found to be stringently constrained by the Super-K data. Especially in the region with relative light DM and heavy dark photon, the Super-K constraint is much stronger than the XENON1T constraint. 
Moreover, we find the PDF effects would reduce the electron recoil rate at the neutrino detectors, thus relaxing the corresponding bounds. The reduction on the rate is $\sim \mathcal{O}(0.1)$ for DM coupling $g^{\prime}=1$ and can reach as small as $10^{-2}$ for DM coupling $g^{\prime}=3$.

We also point out for the first time that the dark photon component in the DM PDFs can induce the dark Compton scattering. 
It can produce a mono-energetic photon in the final states, which can be probed efficiently in future neutrino detectors such as DUNE and JUNO. 
In terms of coupling $\epsilon$, the constraints obtained from the mono-energetic photon at the DUNE and JUNO detectors are around one order of magnitude stronger than those obtained from the electron recoil signal at the Super-K. 

The parameter regions of interest are stringently constrained by simulations of the Bullet Cluster and other galaxy clusters due to strong DM self-interaction. 
Adopting the improved analytical expression for the DM self-scattering cross section, and applying the bound of $\sigma_{\text{self}}/m_{\chi} < 1 ~\text{cm}^{2}/\text{g}$, we find the neutrino detectors can have better sensitivities in the region with relatively heavier dark photon, {\it e.g.} $m_{A^{\prime}} \gtrsim 1$ MeV for $m_{\chi} \lesssim 10^{-4}$ MeV.

Finally, we note that the simplified electron-philic dark photon model adopted in the current work is by no means complete. 
In particular, the DM is overabundant in most regions of interest where the DM is lighter than the dark photon. In this case, the annihilation channel $\chi \chi \to A^{\prime} A^{\prime}$ is kinematically forbidden, and the dark matter dominantly annihilates through the process $\chi \chi \to A^{\prime} \to e^{+}e^{-}$ in the early universe. 
Given the $m_{A^{\prime}} > m_{\chi} \sim \mathcal{O}(1)$ MeV, $g^{\prime} \sim 1$ and $\epsilon \sim 10^{-4}$, the DM relic density can be as high as $10^{4}$. 
However, this problem can be easily solved by either introducing new couplings of $A^{\prime}$ to light SM particles such as neutrinos or introducing new mediator particles such as dark Higgs boson.

\begin{acknowledgments}
This work was supported by the National Natural Science Foundation of China under grant No. 11905149.

\end{acknowledgments}
\appendix

\section{Dark matter scattering off bounded electron}\label{appendix-direct}

For a relativistic CRDM scattering off an electron with effective mass $m_\mathrm{eff}$, $\chi(p_1)+e^{-}(p_2) \rightarrow \chi(k_1)+e^{-}(k_2)$,
the differential cross section can be expressed as
\begin{equation}
\begin{aligned}
d \sigma&=\sum_{nlms}\frac{1}{2 E_{\chi} 2\left(m_{e}-E_{B}^{n l}\right)} \frac{d^{3} k_{2}}{(2 \pi)^{3} 2 E_{k_{2}}} \frac{d^{3} k_{1}}{(2 \pi)^{3} 2 E_{k_{1}}} \frac{d^{3} p_{2}}{(2 \pi)^{3}}(2 \pi)^{4} \delta^{4}\left(p_{1}+p_{2}-k_{1}-k_{2}\right)
\\
&\times \left|i M\left(\bf{p_{1}}, \bf{p_{2}},\bf{k_{1}}, \bf{k_{2}}\right)\right|^{2}\left|\psi_{n l m}\left(\bf{p_{2}}\right)\right|^{2}  ~,~
\end{aligned}
\end{equation}
where the wave function of initial bounded electron in momentum space $\psi_{n l m}\left(\bf{p_{2}}\right) =\chi_{nl}(|{\bf p_2}|)Y_{lm}(\hat{\bf{p}}_{\bf{2}})$ with the normalization $\int d^3 p_2 \left|\psi_{n l m}\left(\bf{p_{2}}\right)\right|^{2}=(2\pi)^3$. $\chi_{nl}(|{\bf p_2}|)$ is radial wave function in momentum space and  $Y_{lm}(\hat{\bf{p}}_{\bf{2}})$ is the spherical harmonic function.

The radial wave function in coordinate space can be obtained by the RHF method. Here, we adopt the expressions provided in reference~\cite{Cao:2020bwd}. Summing over the magnetic quantum number $m$ and the spin $s$ for each $(n,l)$, and using the property of harmonics function $\sum_{m=-l}^{+l}|Y_{lm}(\hat{\bf{p}}_{\bf{2}})|^2=\frac{2l+1}{4\pi}$, we get 

\begin{equation}\label{sigma}
\begin{aligned}
d \sigma &=\sum_{nl}\frac{1}{2 E_{\chi} 2\left(m_{e}-E_{B}^{n l}\right)} \frac{d^{3} k_{2}}{(2 \pi)^{3} 2 E_{k_{2}}} \frac{d^{3} k_{1}}{(2 \pi)^{3} 2 E_{k_{1}}} \frac{d^{3} p_{2}}{(2 \pi)^{3}}(2 \pi)^{4} \delta^{4}\left(p_{1}+p_{2}-k_{1}-k_{2}\right)
\\
&\times \left|i M\left(\bf{p_{1}}, \bf{p_{2}},\bf{k_{1}}, \bf{k_{2}}\right)\right|^{2}|\chi_{n l }(p_2)|^2 \frac{2l+1}{2\pi}~.~
\\
\end{aligned}
\end{equation}

In order to simplify the integration, we apply the variable substitutions as follows
\begin{align}
\text{phase space}&=d^3p_2 d^3k_1 d^3k_2  \delta^4(p_1+p_2-k_1-k_2) \nonumber
\\
&=\frac{1}{4\pi}d\Omega_{p_1}d^3p_2 d^3k_1 d^3 k_2d^3q  \delta(p^0_1+p^0_2-k^0_1-k^0_2)\delta^3(\vec{p}_1-\vec{k}_1-\vec{q})\delta^3(\vec{k}_2-\vec{p}_2-\vec{q}) \nonumber
\\
&=\frac{1}{4\pi}d\Omega_{p_1}d^3p_2  d^3 k_2d^3q  \delta^3(\vec{k}_2-\vec{p}_2-\vec{q}) \delta(p^0_1+p^0_2-k^0_1-k^0_2)~,~  \label{phase}
\end{align} 
where we have introduced an extra integration over solid angle $\Omega_{p_1}$ divided by $4\pi$. 
This corresponds to the global rotation of the reference frame, which does not change the result.

Then, the four-momentum of the DM and the electron can be parameterized by ($p_i$ and $k_i$ denote $|{\bf p_i}|$ and $(|{\bf k_i}|)$ respectively in the following discussion)
\begin{equation}\label{momentum}
\begin{aligned}
&p^\mu_{1}=\left(E_{\chi}, p_{1} \sin \theta_{p_{1}}, 0, p_{1} \cos \theta_{p_{1}}\right) ~,~\\
&p^\mu_{2}=\left(m_{e}-E_{B}^{n l}, p_{2} \sin \theta_{p_{2}} \cos \phi_{p_{2}}, p_{2} \sin \theta_{p_{2}} \sin \phi_{p_{2}}, p_{2} \cos \theta_{p_{2}}\right) ~,~ \\
&k^\mu_{1}=\left(E_{\chi}^{\prime}, p_{1} \sin \theta_{p_{1}}, 0, p_{1} \cos \theta_{p_{1}}-q\right)~,~ \\
&k^\mu_{2}=\left(m_{e}+T_{R}, k_{2} \sin \theta_{k_{2}} \cos \phi_{k_{2}}, k_{2} \sin \theta_{k_{2}} \sin \phi_{k_{2}}, k_{2} \cos \theta_{k_{2}}\right) ~.~
\end{aligned}
\end{equation}
We have chosen the $\vec{q}$ along the $z$-axes and $\vec{p}_1$ in the $x-z$ plane by gauge fixing constraints $\delta(\phi_q-0)$, $\delta(\theta_q-0)$ and $\delta(\phi_{p_1}-0)$. 
In this parameterization, the conversation laws $\delta^3(\vec{k}_2-\vec{p}_2-\vec{q})$ and $\delta(p^0_1+p^0_2-k^0_1-k^0_2)$, as well as the on-shell constraints have not been imposed. However, the conservation law $\delta^3(\vec{p}_1-\vec{k}_1-\vec{q})$ can be satisfied automatically. 

To proceed with the integration, those delta functions are transformed as 

\begin{align}
\delta^{3}  &\left(\vec{k}_{2}-\vec{p}_{2}-\vec{q}\right)  \nonumber\\
&=\frac{1}{k_{2}^{2}} \delta\left(k_{2}-\sqrt{p_{2}^{2}+q^{2}+2 p_{2} q \cos \theta_{p_{2}}}\right) \delta\left(\cos \theta_{k_{2}}-\frac{k_{2}^{2}+q^{2}-p_{2}^{2}}{2 k_{2} q}\right) \delta\left(\phi_{k_{2}}-\phi_{p_{2}}\right) \nonumber \\
&=\frac{1}{k_{2} p_{2} q} \delta\left(\cos \theta_{p_{2}}-\frac{k_{2}^{2}-p_{2}^{2}-q^{2}}{2 p_{2} q}\right) \delta\left(\cos \theta_{k_{2}}-\frac{k_{2}^{2}+q^{2}-p_{2}^{2}}{2 k_{2} q}\right) \delta\left(\phi_{k_{2}}-\phi_{p_{2}}\right) ~,~\label{conservation1}
\end{align}

\begin{align}
 \delta(p^0_1+p^0_2-k^0_1-k^0_2)&=\delta\left(\sqrt{p_{1}^{2}+m_{\chi}^{2}}-\sqrt{m_{\chi}^{2}+p_{1}^{2}+q^{2}-2 p_1 q  \cos \theta_{p_{1}}}-\Delta E_e\right) \nonumber
 \\
 &=\frac{E_{k_{1}}}{p_{1} q} \delta\left(\cos \theta_{p_{1}}-\frac{q^2-\Delta E_e\left(\Delta E_e-2 E_{\chi}\right)}{2 q \sqrt{E_{\chi}^{2}-m_{\chi}^{2}}}\right) ~,~ \label{conservation2}
\end{align}
where $\Delta E_e=T_R+E_B^{nl}$ and $E_{k_1} \equiv k^0_1 \equiv E^\prime_\chi$. 
Using Eq.~\ref{conservation1} and Eq.~\ref{conservation2}, the differential phase space element in Eq.~\ref{phase} becomes

\begin{align}
\text{phase space}&=\frac{1}{4\pi}p_2^2 k_2^2q^2\frac{1}{k_2p_2q}\frac{E_{k_1}}{p_1q}d\phi_{p_1}d\phi_{p_2}d\Omega_q dp_2dk_2dq \nonumber
\\
&=2\pi\frac{E_{k_1}k_2p_2}{p_1}d\phi_{p_2} dp_2dk_2dq \nonumber
\\
&=2\pi\frac{(T_R+m_e)T_RE_{k_1}p_2}{p_1}d\phi_{p_2} dp_2dqd\ln T_R ~,~
\end{align}
where the integration variables $\phi_{p_1}$ and $\Omega_q$ correspond to the global rotation of interaction plane around the $z$-axis and the global rotation of the $\vec{q}$, respectively. 
After considering all momentum conservation constraints and on-shell conditions, only $\phi_{p_2}$, $p_2$, $q$, $T_R$ and $T_\chi(p_1)$ are left as free parameters.

The integration bounds for those parameters are determined by the conditions $-1\leq\cos\theta_{p_2}$, $\cos\theta_{k_2}$, $\cos\theta_{p_1}\leq1$ and $\phi_{p_2}\in(0,2\pi)$. 
The conditions $-1\leq\cos\theta_{p_2},\cos\theta_{k_2}\leq 1$ require $q\in(|k_2-p_2|,|k_2+p_2|)$, and the $-1\leq\cos\theta_{p_1}\leq 1$ gives 
\begin{equation}\label{p1}
p_1>p_1^{\min}=\frac{q}{2}+\frac{\Delta E_e}{2(q^2-\Delta E_e^2)}\sqrt{(q^2-\Delta E_e^2)(q^2-\Delta E_e^2+4m_\chi^2)}, ~~ q>\Delta E_e~.~
\end{equation}
As a result, for given $\ln T_R$, the integration bounds are 
\begin{equation}
\begin{split}
&p_2\in(0,m_e-E^B_{nl})~,~
\\
&q\in(\Delta E_e,+\infty)\cap (|k_2-p_2|,|k_2+p_2|)~,~
\\
&T_\chi\in (T^{\min}_\chi(q),\infty)~,~
\end{split}
\end{equation}
where $T_\chi^{\min}(q)$ is calculated by {\color{blue}Eq.~\ref{p1}} using on-shell condition. Note that for the $p_2$ bound, we also require the electron effective mass to be a real number. 

Furthermore, using the four-momentum parameterizations in Eq.~\ref{momentum}, Eq.~\ref{conservation1} and Eq.~\ref{conservation2}, the matrix element square can be calculated as 
\begin{align}
|iM|^{2}=2 g^{\prime 2} (\epsilon g_{\rm em})^{2}\times \frac{A}{\left(t-m_{A}^{2}\right)^{2}}~,~ \label{amplitude}
\end{align}
where 
\begin{align}
A= &t \left(m_{e}^{2}+2 m_{e} {m}_\mathrm{eff}-m_\mathrm{eff}^{2}+2 m_{\chi}^{2}-4 p^\mu_{1}  k_{2,\mu}\right)-2 m_{\chi}^{2}\left(m_{e}-{m}_\mathrm{eff}\right)^{2} \nonumber \\
&-4 p^\mu_{1}  k_{2,\mu}\left(m_{e}^{2}-{m}_\mathrm{eff}^{2}\right)+8\left(p^\mu_{1}  k_{2,\mu}\right)^{2}+{t^{2}\color{blue}~,}
\end{align} 
with $t = \Delta E^{2}_{e} - q^{2}$ is the four-momentum transfer. 
Finally, we obtain the differential cross section in Eq.~\ref{sigma} and Eq.~\ref{recoil} as 
  
\begin{equation}\label{sigma1}
\begin{aligned}
d \sigma =\frac{2l+1}{16(2\pi)^5}\frac{T_R p_2}{E_\chi(m_e-E_B^{nl})p_1} \left|i M\left(\bf{p_{1}}, \bf{p_{2}},\bf{k_{1}}, \bf{k_{2}}\right)\right|^{2}|\chi_{n l }(p_2)|^2d\phi_{p_2} dp_2dqd\ln T_R~.
\end{aligned}
\end{equation}

\section{Dark matter scattering off free electron}\label{appendix-sk}
In neutrino detectors, the momentum transfer of signal process is much larger than the typical momentum of the initial electron in the atom. So, the initial electron can be approximated to be rest. 

The cross section for a boosted DM/dark photon scattering off a free electron $(\chi/A'(p_1)+e^-(p_2)\rightarrow \chi/\gamma(k_1)+e^-(k_2))$ is given by 
\begin{equation}\label{crossing-sk}
d \sigma =\frac{1}{2 E_{in} 2 m_{e}} \frac{d^{3} k_{2}}{(2 \pi)^{3} 2 E_{k_{2}}} \frac{d^{3} k_{1}}{(2 \pi)^{3} 2 E_{k_{1}}}(2 \pi)^{4} \delta^{4}\left(p_{1}+p_{2}-k_{1}-k_{2}\right)\left|i M\left(p_{1}, p_{2} \rightarrow k_{1}, k_{2}\right)\right|^{2}~,~
\end{equation}
where $E_{in}=E_\chi$ and $E_{A^{\prime}}$ for initial DM and dark photon, respectively. 

For the DM-electron scattering, we choose the momentum transfer $\vec{q}=\vec{k}_2=\vec{p}_1-\vec{k}_1$ alone the $z$-axes and $\vec{p}_1$ in $x$-$z$ plane.
Considering the three momentum conservation $\vec{p}_1+\vec{p}_2-\vec{k}_1-\vec{k}_2 = 0 $, the four-momentum of initial and final state particles can be parameterized as 
\begin{equation}\label{momentum-sk-1}
\begin{aligned}
&p^\mu_{1}=\left(E_{\chi}, p_{1} \sin \theta_{p_{1}}, 0, p_{1} \cos \theta_{p_{1}}\right) \\
&p^\mu_{2}=\left(m_{e}, 0,0,0\right) \\
&k^\mu_{1}=\left(E_{\chi}-T_{R}, p_{1} \sin \theta_{p_{1}}, 0, p_{1} \cos \theta_{p_{1}}-k_2\right) \\
&k^\mu_{2}=\left(m_{e}+T_{R}, 0,0, k_{2} \right)
\end{aligned}
\end{equation}

Then the differential phase space element in Eq.~\ref{crossing-sk} can be simplified to 
\begin{align} 
d^3k_1 d^3k_2 & \delta^{4} \left(p_{1}+p_{2} -k_{1}-k_{2}\right)=\frac{1}{4\pi}d\Omega_{p_1}d^3k_2 \delta\left(p^0_{1}+p^0_{2}-k^0_{1}-k^0_{2}\right) \nonumber
\\
&=\frac{1}{4\pi}d\Omega_{p_1}d^3k_2 \delta\left(E_{\chi}-T_{R}-\sqrt{p_{1}^{2}+k_2^{2}-2 p_{1} k_2 \cos \theta_{p_{1}}+m_{\chi}^{2}}\right) \nonumber
\\
&=\frac{1}{4\pi}d\Omega_{p_1}d^3k_2 \frac{E_{\chi}-T_{R}}{p_{1} k_2} \delta\left(\cos \theta_{p_{1}}-\frac{k_2^{2}-T_{R}\left(T_{R}-2 E_{\chi}\right)}{2 k_2 \sqrt{E_{\chi}^{2}-m_{\chi}^{2}}}\right) \nonumber
\\
&=\frac{1}{4\pi}d\Omega_{p_1}d\Omega_{k_2}k_2^2 dk_2 \frac{E_{\chi}-T_{R}}{p_{1} k_2} \delta\left(\cos \theta_{p_{1}}-\frac{k_2^{2}-T_{R}\left(T_{R}-2 E_{\chi}\right)}{2 k_2 \sqrt{E_{\chi}^{2}-m_{\chi}^{2}}}\right) \nonumber
\\
&=2\pi  \frac{(E_\chi-T_R)(T_R+m_e)T_R}{p_1}d\ln T_R  \label{phase-sk-1}~.~
\end{align}
In the last line, the solid angles $\Omega_{p_1}$ and $\Omega_{k_2}$ which correspond to the global rotation of the reference frame have been integrated out. 

The lower bound of the incident DM kinetic energy for producing a given $T_R$ can be obtained from the condition $-1\leq\cos\theta_{p_1}=\frac{k_2^{2}-T_{R}\left(T_{R}-2 E_{\chi}\right)}{2 k_2 \sqrt{E_{\chi}^{2}-m_{\chi}^{2}}}\leq1$ as shown in the third line of Eq.~\ref{phase-sk-1}, which is
\begin{equation}\label{range-sk-1}
\begin{split}
T_\chi>T_\chi^{\min}&=\left(\frac{T_{R}}{2}-m_{\chi}\right)\left[1 \pm \sqrt{1+\frac{2 T_{R}}{m_{e}} \frac{\left(m_{e}+m_{\chi}\right)^{2}}{\left(2 m_{\chi}-T_{R}\right)^{2}}}\right]~,~
\end{split}
\end{equation}
where the +(-) sign is used for $T_R>2m_\chi~(T_R<2m_\chi)$. 

For dark photon scattering off electron $A'(p_1)+e^-(p_2)\rightarrow \gamma(k_1)+e^-(k_2)$, concerning the electron recoil energy, we choose  $\vec{q}=\vec{k}_2=\vec{p}_1-\vec{k}_1$ alone $z$-axes, and $\vec{p}_1$ in $x$-$z$ plane as the same with the DM scattering. 
As a result, the  parameterizations of the four-momentum of initial and final state particles are similar as those shown in Eq.~\ref{momentum-sk-1}, with the subscript $\chi$ replaced by $A^{\prime}$. And the differential phase space element is given by 
\begin{align}
d^3k_1 d^3k_2 & \delta^{4}\left(p_{1}+p_{2}-k_{1}-k_{2}\right)=\frac{1}{4\pi}d\Omega_{p_1}d^3k_2 \delta\left(p^0_{1}+p^0_{2}-k^0_{1}-k^0_{2}\right) \nonumber
\\
&=\frac{1}{4\pi}d\Omega_{p_1}d^3k_2 \delta\left(E_{A^{\prime}}-T_{R}-\sqrt{p_{1}^{2}+k_2^{2}-2 p_{1} k_2 \cos \theta_{p_{1}}}\right) \nonumber
\\
&=\frac{1}{4\pi}d\Omega_{p_1}d^3k_2 \frac{E_{A^{\prime}}-T_{R}}{p_{1} k_2} \delta\left(\cos \theta_{p_{1}}-\frac{-m_{A^{\prime}}^{2}+k_2^{2}-T_{R}\left(T_{R}-2 E_{A^{\prime}}\right)}{2 k_2 \sqrt{E_{A^{\prime}}^{2}-m_{A^{\prime}}^{2}}}\right) \nonumber
\\
&=\frac{1}{4\pi}d\Omega_{p_1}d\Omega_{k_2}k_2^2dk_2 \frac{E_{A^{\prime}}-T_{R}}{p_{1} k_2} \delta\left(\cos \theta_{p_{1}}-\frac{-m_{A^{\prime}}^{2}+k_2^{2}-T_{R}\left(T_{R}-2 E_{A^{\prime}}\right)}{2 k_2 \sqrt{E_{A^{\prime}}^{2}-m_{A^{\prime}}^{2}}}\right) \nonumber
\\
&=2\pi  \frac{(E_\chi-T_R)(T_R+m_e)T_R}{p_1}d\ln T_R~,~ \label{phase-sk-2}
\end{align}
Similarly, the lower bound of the incident dark photon kinetic energy for producing a given $T_R$ can be obtained from the condition $-1\leq\cos\theta_{p_1}=\frac{-m_{A^{\prime}}^{2}+k_2^{2}-T_{R}\left(T_{R}-2 E_{A^{\prime}}\right)}{2 k_2 \sqrt{E_{A^{\prime}}^{2}-m_{A^{\prime}}^{2}}}\leq1$ as shown in the fourth line of Eq.~\ref{phase-sk-2}:
\begin{align}
T_{A^{\prime}}> & T_{A^{\prime}}^{\min}=\frac{-m_{A^{\prime}}^{2}-4 m_{A^{\prime}} m_{e}
+2 m_{e} T_{R}}{4 m_{e}} \nonumber
\\
&+\frac{1}{4} \sqrt{\frac{2 m_{A^{\prime}}^{4} m_{e}+m_{A^{\prime}}^{4} T_{R}+8 m_{A^{\prime}}^{2} m_{e}^{2} T_{R}+4 m_{A^{\prime}}^{2} m_{e} T_{R}^{2}+8 m_{e}^{3} T_{R}^{2}+4 m_{e}^{2} T_{R}^{3}}{m_{e}^{2} T_{R}}}~.~ \label{range-sk-2}
\end{align}

Finally, for dark photon scattering off electron $A^{\prime}(p_1)+e^-(p_2)\rightarrow \gamma(k_1)+e^-(k_2)$ while concerning the outgoing photon energy, the same reference frame is chosen, {\it i.e.,} $\vec{q}$ along $z$-axes and $\vec{p}_1$ in $x$-$z$ plane. 
Considering the three-momentum conservation $\vec{q}-\vec{k}_2 = 0 $, the four-momentum of initial and final state particles can be parameterized as 
\begin{equation}
\begin{aligned}
p_{1} &=\left(E_{A^{\prime}}, p_{1} \sin \theta_{p_{1}}, 0, p_{1} \cos \theta_{p_{1}}\right) \\
p_{2} &=\left(m_{e}, 0,0,0\right) \\
k_{1} &=\left(E_{\gamma}, E_{\gamma} \sin \theta_{k_{1}} \cos \phi_{k_{1}}, E_{\gamma} \sin \theta_{k_{1}} \sin \phi_{k_{1}}, E_{\gamma} \cos \theta_{k_{1}}\right) \\
k_{2} &=\left(E_{A^{\prime}}+m_{e}-E_{\gamma}, 0,0, q\right)
\end{aligned}
\end{equation}
where $p_1=\sqrt{E_{A^{\prime}}-m_{A^{\prime}}^2}$.
And the differential phase space element is expressed by
\begin{equation}\label{phase-sk-3.0}
\begin{aligned}
d^{3} k_{1} d^{3} k_{2} & \delta\left(p_{1}^{0}+p_{2}^{0}-k_{1}^{0}-k_{2}^{0}\right) \delta^{3}\left(\vec{p}_{1}+\vec{p}_{2}-\vec{k}_{1}-\vec{k}_{2}\right) \\
=&\frac{d\Omega_{p_1}}{4\pi}d^{3} k_{1} d^{3} k_{2} d^{3} q \delta\left(p_{1}^{0}+p_{2}^{0}-k_{1}^{0}-k_{2}^{0}\right) \delta^{3}\left(\vec{p}_{1}-\vec{k}_{1}-\vec{q}\right) \delta^{3}\left(\vec{k}_{2}-\vec{q}\right) \\
=& \frac{d\Omega_{p_1}}{4\pi}d^{3} k_{1} d^{3} q \delta\left(p_{1}^{0}+p_{2}^{0}-k_{1}^{0}-k_{2}^{0}\right) \delta^{3}\left(\vec{p}_{1}-\vec{k}_{1}-\vec{q}\right)~,~
\end{aligned}
\end{equation}
where the delta functions can be transformed as
\begin{align}\label{conservation-sk-3}
 \delta^3 & \left(\vec{p}_{1}-\vec{k}_{1}-\vec{q}\right)  \nonumber\\
 &=\frac{1}{p_{1}^{2}} \delta\left(p_{1}-\sqrt{E_{\gamma}^{2}+q^{2}+2 q E_{\gamma} \cos \theta_{k_{1}}}\right) \delta\left(\cos \theta_{p_{1}}-\frac{p_{1}^{2}+q^{2}-E_{\gamma}^{2}}{2 p_{1} q}\right) \delta\left(0-\phi_{k_{1}}\right)  \nonumber
 \\ 
 &=\frac{1}{p_{1} q E_{\gamma}} \delta\left(\cos \theta_{k_{1}}-\frac{p_{1}^{2}-E_{\gamma}^{2}-q^{2}}{2 q E_{\gamma}}\right) \delta\left(\cos \theta_{p_{1}}-\frac{p_{1}^{2}+q^{2}-E_{\gamma}^{2}}{2 p_{1} q}\right) \delta\left(0-\phi_{k_{1}}\right) ~,~  \nonumber\\
 \delta (&p_1^0+p_2^0-  k_1^0-k_2^0)=\delta\left(\left(E_{A^{\prime}}+m_{e}-E_{\gamma}\right)-\sqrt{q^{2}+m_{e}^{2}}\right) \nonumber
 \\
 &=\frac{E_{A^{\prime}}+m_{e}-E_{\gamma}}{q} \delta\left(q-\sqrt{\left(E_{A^{\prime}}+m_{e}-E_{\gamma}\right)^{2}-m_{e}^{2}}\right)
\end{align}
So, the Eq.~\ref{phase-sk-3.0} can be further simplified to 
\begin{equation}\label{phase-sk-3}
\begin{aligned}
\mathrm{phase ~space ~element}=&\frac{E_\gamma^2(E_{A^{\prime}}+m_{e}-E_{\gamma})}{4\pi p_1} d \Omega_{q} d \phi_{p_{1}}  d \ln E_\gamma
\\
=&2\pi\frac{E_\gamma^2(E_{A^{\prime}}+m_{e}-E_{\gamma})}{ p_1}  d \ln E_\gamma
\end{aligned}
\end{equation}
The lower and upper bounds of the incident dark photon energy for producing a given photon energy $E_\gamma$ can be obtained from the condition $-1\leq\cos\theta_{k_1},\cos\theta_{p_1}\leq 1$ and $q \in \mathcal{R}$ as shown in the delta functions of Eq.~\ref{conservation-sk-3}, which are 
\begin{equation}\label{range-sk-3}
\begin{split}
E_{A^{\prime}} \in\begin{cases}
&(E_{A^{\prime}}^-,E_{A^{\prime}}^+) ~~~~m_{A^{\prime}}<m_e ~\text{and}~\frac{-m_{A^{\prime}}^2+2m_{A^{\prime}} m_e}{2m_e}<E_\gamma<\frac{m_e}{2}
\\
&(E_{A^{\prime}}^+, \infty)~~~~E_\gamma>\frac{m_e}{2}
\end{cases}
\end{split}~,~
\end{equation}
where $E_{A^{\prime}}^-$ and $E_{A^{\prime}}^+$ are defined as the following
\begin{align}
E_{A^{\prime}}^{\pm}= &\frac{(E_\gamma-m_e)(-m_{A^{\prime}}^2+2E_\gamma m_e)}{(2E_\gamma-m_e)2m_e} \nonumber \\
& \pm\frac{E_\gamma}{2m_e}\sqrt{\frac{(m_{A^{\prime}}^2+2E_\gamma m_e-2m_{A^{\prime}} m_e)(m_{A^{\prime}}^2+2(E_\gamma+m_{A^{\prime}})m_e)}{(-2E_\gamma+m_e)^2}}~.~
\end{align} 
Note that in our case, we only focus on the branch with $E_\gamma>\frac{m_e}{2}$. 

The only unknown ingredient of the cross section in Eq.~\ref{crossing-sk} is the matrix element square. The matrix element squares for the DM-electron scattering with respect to the recoiled electron kinetic energy $|iM^e_{\chi e}|^{2}$, the dark photon-electron scattering with respect to the recoiled electron kinetic energy $|iM^e_{A^{\prime} e}|^{2}$ as well as the dark photon-electron scattering with respect to the outgoing photon energy $|iM^\gamma_{A^{\prime} e}|^{2}$ can be calculated as 
\begin{align}
|iM^e_{\chi e}|^{2}&=2g^{\prime 2}(\epsilon g_{\text{em}})^2\frac{2(s(t-2m_e^2-2m_\chi^2)+(m_e^2+m_\chi^2)^2+s^2)+t^2}{(t-m_{A^{\prime}}^2)^2}~,~
\\
|iM^e_{A^{\prime} e}|^{2}=&\frac{4}{3}(\epsilon g_{\text{em}})^2 g_{\text{em}}^2\frac{1}{(m_e^2-s)^2(m_e^2-u)^2} \times \{-2m_{A^{\prime}}^4(m_e^2-s)(m_e^2-u) \nonumber
\\
&+2m_{A^{\prime}}^2(m_e^4(s+u)-4m_e^2su+su(s+u))+6m_e^8-m_e^4(3s^2+14su+3u^2) \nonumber
\\
&+m_e^2(s+u)(s^2+6su+u^2)-su(s^2+u^2)\}~,~
\\
|iM^\gamma_{A^{\prime} e}|^{2}&=|iM^e_{A^{\prime} e}|^{2}~,~
\end{align}
where the Mandelstam variables $s$, $t$, $u$ for each process are defined as 
\begin{equation}\label{sk-amplitude}
\begin{split}
&iM^e_{\chi e}:
\begin{cases}
&s=(m_\chi+m_e)^2+2m_e T_\chi
\\
&t=-2m_e T_R
\end{cases}~,~
\\
&iM^e_{A^{\prime} e}:
\begin{cases}
&s=(m_{A^{\prime}}+m_e)^2+2m_e T_{A^{\prime}}
\\
&u=m_e^2-2m_e(m_{A^{\prime}}+T_{A^{\prime}}-T_R)
\end{cases}~,~
\\
&iM^\gamma_{A^{\prime} e}:
\begin{cases}
&s=(m_{A^{\prime}}+m_e)^2+2m_e T_{A^{\prime}}
\\
&u=m_e^2-2m_eE_\gamma
\end{cases}~.~
\end{split}
\end{equation}

Combining the matrix elements with the phase space elements, we obtain the differential cross sections for all of the processes
\begin{equation}\label{sk-crossing}
\begin{split}
\frac{d\sigma^e_{\chi e}}{d\ln T_R}&=\frac{1}{32\pi}\frac{T_R}{p_1E_\chi m_e }|iM_{\chi e}^e|^2 ~,~
\\
\frac{d\sigma^e_{A^{\prime} e}}{d\ln E_\gamma}&=\frac{1}{32\pi}\frac{T_R}{p_1E_{A^{\prime}} m_e}|iM_{A^{\prime} e}^e|^2 ~,~
\\
\frac{d\sigma^\gamma_{A^{\prime} e}}{d\ln E_\gamma}&=\frac{1}{32\pi}\frac{E_\gamma}{p_1E_{A^{\prime}} m_e}|iM^{\gamma}_{A^{\prime} e}|^2~.~
\end{split}
\end{equation}

\bibliographystyle{jhep}
\bibliography{BoostDM}
\end{document}